\documentclass[showkeys,prd,nofootinbib,floatfix,preprintnumbers]{revtex4-1}

\usepackage{amsfonts,amsmath,amssymb} 
\usepackage{graphicx,graphics,color}
\usepackage{gensymb}
\usepackage{longtable}
\usepackage{bbding}
\usepackage{hyperref}

\begin{document}


\title{Conformally and Disformally Coupled Vector Field Models of Dark Energy}

\author{Gabriel G\'omez}

\affiliation{Departamento de F\'isica, Universidad de Santiago de Chile,\\Avenida V\'ictor Jara 3493, Estaci\'on Central, 9170124, Santiago, Chile}

\date{\today}

\begin{abstract}
Scalar fields coupled to dark matter by conformal or disformal transformations give rise to a general class of scalar-tensor theories which leads to a rich phenomenology in a cosmological setting. 
While this possibility has been studied comprehensively in the literature for scalar fields, the vector case has been hardly treated. We build hence models based on vector fields conformally and disformally coupled to dark matter and derive explicitly the general covariant form of the interaction term in an  independent way of the gravity theory, whereby this result can be applied to general vector-tensor theories. For concreteness, the standard Proca theory with a vector exponential potential is taken to describe the vector-tensor sector, and some specific coupling functions are assumed to study the cosmological background dynamics by dynamical system techniques. As a first examination about instabilities issues, we derive general conditions to avoid classical instabilities in a more general setup of the theory. Interestingly, 
despite choosing such a minimalist form for the underlying theory, the parameter space is considerably enriched compared to the uncoupled case due to the novel interactions, leading to new branches of solutions for the vector equation of motion. Thus, different trajectories can exist in phase space depending on the coupling parameters associated to the conformal and disformal functions. From here, new emerging vector-dark matter scaling solutions, and renewed stable attractor points are found to drive the late-time accelerated expansion of the universe. Numerical calculations are performed as well to investigate more quantitatively the impact of the conformal and disformal couplings on the cosmological background evolution. These effects depend essentially on the strength of the coupling parameters and, in some specific cases, on their associated signs. In all the cases studied we find that the coupling of the vector field to dark matter can affect significantly the cosmological dynamics during different stages of the evolution of the universe.



\end{abstract}

\pacs{Valid PACS appear here}
\keywords{}
\maketitle


\section{Introduction}

The golden age of cosmology, referred commonly to unprecedented progress in observational cosmology, accompanied by impressive development in theoretical grounds have shaped firmly our understanding of the Universe. Specifically, 
James Peebles, awarded a Novel Prize in physics in 2019 for theoretical development in physical cosmology, has contributed, among other prominent cosmologists, to the basis of our contemporary conception about the universe \cite{Peebles:1994xt}. One of the
most fascinating features of the universe is that most of its energy content, that is around $70\%$ according to the cosmic radiation background analysis \cite{WMAP:2012nax,Planck:2018vyg}, is in the form of dark energy, a mysterious repulsive force pushing galaxies apart \cite{Riess:1998cb,Schmidt:1998ys,Perlmutter:1998np} and whose nature is still unknown. On the other hand, high-precision measurements in cosmology  such as anisotropies in the cosmic microwave background temperature and polarization fields, weak lensing, galaxy clustering, standard candles and baryon acoustic oscillations \cite{Alam:2016hwk,Planck:2018vyg,Jones:2017udy,DES:2021wwk,SNLS:2010pgl,SupernovaCosmologyProject:2011ycw,SNLS:2011lii,ACT:2020gnv,2dFGRS:2005yhx,SDSS:2006lmn,BOSS:2016wmc,HSC:2018mrq,Heymans:2020gsg,HSC:2018mrq} and, recently, the direct detection of gravitational waves by LIGO and Virgo collaborations \cite{LIGOScientific:2017zic,TheLIGOScientific:2017qsa} have been used as the major observational discriminators of gravity theories that attempt to describe consistently the 
current accelerated expansion of the universe. This is indeed of great concern today because of the emergent tensions in the $\Lambda$CDM cosmological model when confronting with observations. These discrepancies are specifically due to a lower rate for the cosmic growth derived from observations of the redshift-space distorsion \cite{Macaulay:2013swa} and cluster counts \cite{Battye:2014qga,Alam:2016hwk,Abbott:2017wau}, and a lack of conciliation between early and late measurements which has been referred to, ever since, as the Hubble tension  \cite{Freedman:2017yms,Verde:2019ivm}.

A fundamental and consensual description of the underlying physical mechanism for the agent driving the current accelerated expansion is still lacking. Although the simplest explanation, within the $\Lambda$CDM cosmological model \cite{Planck:2018vyg}, is identifying the cosmological constant as the agent responsible for the accelerated expansion, it leads to a tremendous discrepancy (of around $120$ orders of magnitude due to zero-point contributions to vacuum fluctuations) when compared with its observed value \cite{Weinberg:1988cp,Amendola:2015ksp}. One way to evade (not to solve) this problem, that must be treated indeed in any alternative scenario to the accelerated expansion, is to resort to some mechanisms in which the cosmological constant vanishes or becomes negligible compared to present cosmological energy density \cite{Amendola:2015ksp}. Though it is argued sometimes that those mechanisms are present in dynamical dark energy models,   generally rooted within higher-dimensional theories, the truth is that there is not a clear solution to tackle this problem and one must assume simply a vanishing cosmological constant. Traditionally canonical (quintessence) \cite{Wetterich:1987fm,Ratra:1987rm} and non-canonical (k-essence) scalar fields \cite{ArmendarizPicon:1999rj,ArmendarizPicon:2000dh,ArmendarizPicon:2000ah} are identified as dynamical dark energy \cite{Peebles:2002gy}. However, one can go beyond these conventional approaches by modifications of the geometric sector of Einstein gravity by breaking its fundamental assumptions \cite{Lovelock:1971yv,Horava:2009uw}, or by including extra fields non-minimally coupled to gravity  \cite{Horndeski:1974wa,Horndeski:1976gi,Nicolis:2008in,Deffayet:2011gz,Kobayashi:2011nu,Heisenberg:2014rta,Allys:2016jaq,BeltranJimenez:2016rff,GallegoCadavid:2020dho}. 

An intermediate approach to account also for the aforementioned discrepancies in the $\Lambda$CDM cosmological model is to assume phenomenological interactions between dark mater and dark energy \cite{DiValentino:2017iww,DiValentino:2019jae}. This idea has been extensively explored in the literature by taking at hand a wide variety of interaction types (see e.g. \cite{Bahamonde:2017ize,Wang:2016lxa} and references therein) but missing, in most of the cases,  justification from the theoretical point of view\footnote{An interesting proposal comes directly from the quantum field theory of Einstein-Cartan gravity \cite{Begue:2017lcw}.}. A more grounded way to account for the interactions is to build interactions at the level of the actions by, for instance, conformal and disformal transformations; the latter introduced originally by Bekenstein to relate geometries of the same gravitational theory \cite{Bekenstein:1992pj}. This possibility has been exploited extensively in the context of scalar-tensor theories
\cite{Amendola:1999er,Amendola:1999qq,Koivisto:2008ak,vandeBruck:2015ida,Zumalacarregui:2010wj,Zumalacarregui:2012us,vandeBruck:2013yxa,vandeBruck:2016jgg,vandeBruck:2016hpz,Teixeira:2019hil,Teixeira:2019tfi,Chibana:2019jrf,Thipaksorn:2022yul}
but only partially in vector-tensor theories\footnote{Interplay between both sectors can also result in an interesting cosmological setting \cite{Thorsrud:2012mu,Koivisto:2014gia,Gomez:2021jbo}.} (see e.g. \cite{Gomez:2020sfz}) what motivates us, therefore, to investigate such a possibility from a consistent and comprehensive framework as will be discussed below. Alternatively, conformal and disformal transformations have become a complementary mathematical tool in the understanding of the structure of generalized scalar-tensor theories \cite{Bettoni:2013diz,Achour:2016rkg,Zumalacarregui:2013pma,Gleyzes:2014qga,deRham:2010ik,Ezquiaga:2018btd,Zumalacarregui:2012us}. Though there does not exist a guiding principle to build interactions from this approach,  there is not a physical reason either, unless some symmetry principle or fundamental law are imposed, to think that the metric associated to dark matter is exactly equal to that of the gravity sector. It is reasonable to think that this statement can be also valid for theories when fields are non-minimally coupled to gravity. It should be stressed that the coupling between different sectors appears naturally in the context of higher dimensional theories and theories of massive gravity \cite{deRham:2010kj}, and emerges generically in brane-world scenarios where matter fields reside on a hidden moving brane \cite{Koivisto:2013fta}. In scalar-vector-tensor theories, like TeVeS, the two metrics involved are related by disformal transformations that relate non-trivially the fields involved \cite{Bekenstein:2004ne}.

On the other hand, models involving vector or gauge fields have a long standing history in cosmological contexts \cite{Ford:1989me, Jacobson:2000xp,Dimopoulos:2006ms, Ackerman:2007nb,Dimopoulos:2008yv,Golovnev:2008cf,EspositoFarese:2009aj,Dimopoulos:2011ws}. Although most of the early works have focused on the role of vector fields during the inflationary period \cite{Emami:2016ldl,Maleknejad:2012fw,Soda:2012zm,Dimastrogiovanni:2010sm,Bamba:2008xa,Garnica:2021fuu}, some authors have also been interested in the possibility of driving the late-time evolution of the universe, either when they are coupled minimally to gravity  \cite{ArmendarizPicon:2004pm, Koivisto:2007bp, Koivisto:2008ig,Mehrabi:2015lfa,Guarnizo:2020pkj,Gomez:2020sfz} or in more general theories of gravity when the vector field plays the role of a new degree of freedom of gravity \cite{Tasinato:2014eka, DeFelice:2016yws,Rodriguez:2017wkg,Nakamura:2017dnf,Geng:2021jso,Heisenberg:2020xak}. It is important to mention that some significant progresses in the construction of coupled vector dark energy models have been done recently, 
following different approaches to the one we are interested here\footnote{Some of these works rely on phenomenological couplings to account for the dark sector interaction.} \cite{Zhao:2008tk,Koivisto:2012xm,Ngampitipan:2011se,Yao:2017enb,Wei:2006tn,Landim:2016dxh,Koivisto:2008xf,Nakamura:2019phn,Yao:2020pji}. Nevertheless, closer to the spirit of the present paper it was proposed a new class of conformally coupled dark energy model based on (space-like) multi-vector fields through a conformal transformation \cite{Gomez:2020sfz}. One might wonder then whether more general interactions than the one presented in \cite{Gomez:2020sfz} can be built from disformal transformations following the same mathematical approach as in coupled scalar fields models of dark energy. This is the main problem we want to deal with in this paper. 

Thus, motivated by the salient role of vector fields in cosmology along with the phenomenological perspectives of coupled dark energy models, we propose in this paper to build interactions between the gravitational sector, identified by the vector field, and dark matter via a vector disformal transformation which relates the geometry of both sectors. As a concrete example to see how the resulting interactions operate at the background level, the standard Proca theory, and a vector exponential potential assumed to describe dark energy, are taken to describe the gravitational sector of the model as a proof of concept. On the other hand, the conformal and disformal couplings are assumed to be functional of the (vector) fields only to guarantee safely second-order field equations and, thus, to avoid the presence of Ostrogradski instabilities at this stage. As a general result, the derived interaction term is quite independent of the gravity theory and can be applied to more general vector-tensor theories as the Generalized Proca theory \cite{Heisenberg:2014rta,Allys:2015sht,BeltranJimenez:2016rff,Allys:2016jaq}. Some particular choices of the coupling functions are considered by concreteness in order to investigate the background evolution by dynamical system analysis. From here, new critical points arise, enriching considerably the parameter space in comparison to the uncoupled case. In all the cases studied, the effect of the associated coupling parameters are quite significant in the background evolution of the universe. Thus, these results constitute an archetype towards building more general models of coupled vector dark energy involving, for instance, first-order derivatives of the vector field within more general vector-tensor theories that can account, among other phenomenological aspects, for the accelerated expansion. These results also suggest that the use of observational data at different redshifts (depending on the coupling type) is imperative to put constrains on the model parameters, along with the ones derived here from purely theoretical grounds, aiming at ameliorating the current discrepancies in the $\Lambda$CDM cosmological model.

The content of this paper is structured as follows. In section \ref{sec:2}, the covariant form of the interaction term is derived  assuming a field-dependent disformal transformation but independent of the gravity theory. In section \ref{sec:3}, the evolution equations that govern the background dynamics are found for a particular model. In section \ref{sec:4}, dynamical system techniques are implemented to investigate the cosmological background dynamics for particular choices of the coupling functions. Complementary to this study, we apply some numerical methods in section \ref{sec:5} to assess more quantitatively the effect of the coupling parameters in the evolution of the universe. Finally, a general discussion of the results found and some perspectives of this work are presented in section \ref{sec:6}.

\section{Vector Disformal coupling to dark matter}\label{sec:2}

We start with a general class of vector-tensor theories minimally coupled to gravity but allowing higher-order derivatives  self-interaction\footnote{A general vector-tensor theory
that contains up to two derivatives with respect to metric and vector field has been built as an extension of a massive vector theory in curved space-time \cite{Kimura:2016rzw,GallegoCadavid:2021ljh}.}, through the gauge-invariant term $Y=-\frac{1}{4}F_{\mu\nu}F^{\mu\nu}$, with $F_{\mu\nu} \equiv \nabla_{\mu}A_{\nu}-\nabla_{\nu}A_{\nu}$, and an explicit symmetry breaking through the quantity $X=-\frac{1}{2}g^{\mu\nu}A_{\mu}A_{\nu}$, and a cold dark matter Lagrangian coupled (non-trivially) to the gravitational sector. Accordingly, the action can be expressed in the Einstein frame as 
\begin{equation}
   \mathcal{S}=\int d^{4}x\left[\sqrt{-g} \left(\frac{M_{p}^2}{2}R +\mathcal{L}_{A}(X,Y)\right)+\sqrt{-\bar{g}} \bar{\mathcal{L}}_{c}[\bar{g}_{\mu\nu},\psi_{c}]\right],\label{sec2:eqn1}
\end{equation}
where $M_p$ is the reduced Planck mass, $R$ is the Ricci scalar and $\psi_{c}$ is the matter field. The dark matter Lagrangian follows, therefore, geodesics defined by the barred metric $\bar{g}_{\mu\nu}$ which differ from to the ones described by the gravitational sector $g_{\mu\nu}$. Both metrics are related by a  vector disformal transformation of the form\footnote{This kind of vector disformal transformation was firstly introduced in the literature to build general self-interactions of the vector field in a Minkowski background at the desired order \cite{BeltranJimenez:2016rff}.} \cite{BeltranJimenez:2016rff,Kimura:2016rzw}
\begin{equation}\bar{g}_{\mu \nu}=C(X)g_{\mu \nu}+B(X) A_{\mu} A_{\nu}.\label{sec2:eqn2}
\end{equation}
The barred inverse metric is given by
\begin{equation}\bar{g}^{\mu \nu}=\frac{1}{C}\left(g^{\mu \nu}-\frac{B}{C-2 B X} A^{\mu} A^{\nu}\right),\label{sec2:eqn3}
\end{equation}
and the coupling functions $C(X)$ and $B(X)$ are arbitrary vector field dependent functions assumed to depend, as the main theoretical assumption, on the mass-like term $X$ only, i.e., on the field itself and not on its derivatives. It is also possible to include here dependence of (powers of) the Maxwell term $Y$ (and its dual) but it may lead to higher order equations of motions. This possibility is then excluded in the present study in order to avoid safely Ostrogradski instabilities at this stage of the construction\footnote{Note that although the inclusion of higher derivative terms lead inevitably to the propagation of unwanted degree of freedoms, it is possible to integrated them out by a Hamiltonian constraint. This approach was used particularly when general disformal transformations involving powers of the field strength tensor $F_{\mu\nu}$ are used to build non-linear extensions of the Einstein-Maxwell theory \cite{Gumrukcuoglu:2019ebp}.}. In the context of the Generalized Proca theory, other pieces beyond the $\mathcal{L}_{2}$ (identified here simply as $\mathcal{L}_{A}$), as the $\mathcal{L}_{3}$, which is absent in the non-Abelian version of the theory \cite{Allys:2016kbq, GallegoCadavid:2020dho,Gomez:2019tbj}, can be included for generality. The latter however introduces additional degrees of freedom that can, in turn, lead to overcloud the already known conditions for the avoidance of Laplacian and ghost instabilities \cite{DeFelice:2016yws} due to the non-trivial coupling to dark matter. 
By varying the action  with respect to the metric, the gravitational field equations in the Einstein Frame yield
\begin{equation}
    \frac{M_{p}^{2}}{2}G_{\mu\nu}=T^{(A)}_{\mu\nu}+T^{(c)}_{\mu\nu}+T^{(i)}_{\mu\nu}.\label{sec2:eqn3a}
\end{equation}
where the energy momentum tensor of each component are defined respectively as 
\begin{equation}
T_{(A)}^{\mu \nu}=\frac{2}{\sqrt{-g}} \frac{\delta\left(\sqrt{-g} \mathcal{L}_{A}\right)}{\delta g_{\mu \nu}}, \quad T_{(\mathrm{c})}^{\mu \nu}=\frac{2}{\sqrt{-g}} \frac{\delta\left(\sqrt{-\bar{g}} \overline{\mathcal{L}}_{\mathrm{c}}\right)}{\delta g_{\mu \nu}},\quad T_{(i)}^{\mu \nu}=\frac{2}{\sqrt{-g}} \frac{\delta\left(\sqrt{-g} \mathcal{L}_{i}\right)}{\delta g_{\mu \nu}}.\label{sec2:eqn4} \end{equation}
where the  index $i=\rm r,b$ stands for the radiation and baryons components, respectively, which evolve in the standard manner since they are minimally coupled to the gravitational sector\footnote{They may be affected however in an indirect way by the coupling since gravity acts as a messenger between all the components.}. In order to relate the energy momentum tensor for dark matter in the Jordan (barred) and Einstein (unbarred) frames, it is necessary to find the relation between the determinant of the barred and unbarred metrics
\begin{equation}\sqrt{-\bar{g}}=\sqrt{-g} \sqrt{C^{3}(C -2 B X)}.\label{sec2:eqn5}
\end{equation}
From all the above, the energy momentum tensor of dark matter in both frames follows the relation
\begin{equation}
T_{(\mathrm{c})}^{\mu \nu}=\sqrt{\frac{\bar{g}}{g}} \frac{\partial \bar{g}_{\alpha \beta}}{\partial g_{\mu \nu}} \bar{T}_{(\mathrm{c})}^{\alpha \beta},\label{sec2:eqn6}
\end{equation}
where the energy momentum tensor in the barred frame has been defined as
\begin{equation}
\bar{T}_{\mathrm{(c)}}^{\alpha \beta}=\frac{2}{\sqrt{-\bar{g}}} \frac{\delta\left(\sqrt{-\bar{g}} \mathcal{L}_{c}\right)}{\delta \bar{g}_{\alpha \beta}}.\label{sec2:eqn7}
\end{equation}
Differentiating explicitly eqn.~(\ref{sec2:eqn2}) with respect to the unbarred metric gives the Jacobian of the transformation required in eqn.~(\ref{sec2:eqn6}) to transform the energy momentum tensor from one frame to another. This is showed in the appendix along with other useful relations. The explicit transformation is
\begin{equation}T_{(\mathrm{c})}^{\mu \nu}=\sqrt{\frac{\bar{g}}{g}}\left[C \bar{T}_{(\mathrm{c})}^{\mu \nu}+\frac{1}{2}A^{\mu} A^{\nu}\left(C_{,X} g_{\alpha \beta}+B_{,X} A_{\alpha} A_{\beta}\right) \bar{T}_{(\mathrm{c})}^{\alpha \beta} \right].\label{sec2:eqn8}
\end{equation}
Here the subscripts $X$ (and $Y$ to be used later) represents  derivatives with respect to the mass term (and its kinetic term). Thus, in order to preserve the isotropy of the background we choose the  temporal component of the vector field only. It leads, as a result, to have a presureless fluid in both frames. Nevertheless, we can start from a situation in which the fluid is defined presureless in the unbarred metric but once one assumes, for instance, a non-vanishing spatial configuration for the vector field, taking  three copies of canonical Maxwell fields to be also consistent with the background properties, an effective pressure can arise in the Jordan frame due to the non-minimal coupling between matter and (spatial) vector fields. This feature is present in the conformally coupled Multi-Proca vector dark energy model \cite{Gomez:2020sfz} due to the second term of eqn.~(\ref{sec2:eqn8}). This does not happen however for the scalar field case where the dark energy fluid is always presureless in both frames.

After extremizing the action with respect to the vector field, one gets the relation  $\frac{\delta\mathcal{L}_{A}} {\delta A_{\alpha}}=-\frac{1}{\sqrt{-g}} \frac{\delta(\sqrt{-\bar{g}}\overline{\mathcal{L}}_{c})} {\delta A_{\alpha}}$, which is the Euler-Lagrange equation sourced by the coupling\footnote{Notice that $Q^{\alpha}$ can be thought of as the component of an electric current by analogy with electromagnetism.} vector field to dark matter $Q^{\alpha}$
\begin{equation}
\frac{\partial \mathcal{L}_{A}}{\partial A_{\alpha}}-\nabla_{\beta} \frac{\partial \mathcal{L}_{A}}{\partial\left(\nabla_{\beta} A_{\alpha}\right)}=Q^{\alpha}.\label{sec2:eqn9}
\end{equation}
It can be written in a more explicit and compact way given the dependence of the vector Lagrangian  as
\begin{equation}
 \mathcal{L}_{A,Y} \nabla_{\beta} F^{\alpha\beta}+ \mathcal{L}_{A,X} A^{\alpha}+ \mathcal{M}_{\beta} F^{\alpha\beta}=Q^{\alpha}, \label{sec2:eqn10}
\end{equation}
with $\mathcal{M}_{\beta}= \mathcal{L}_{A,XX} A^{\nu}\nabla_{\beta}A_{\nu}+ \mathcal{L}_{A,YY}F^{\rho\nu}\nabla_{\beta}\nabla_{\nu}A_{\rho}$ and $Q^{\alpha}$ has been defined as
\begin{equation}
Q^{\alpha}=-\frac{1}{\sqrt{-g}} \frac{\delta(\sqrt{-\bar{g}}\overline{\mathcal{L}}_{c})}{\delta A_{\alpha}}=-\frac{1}{\sqrt{-g}}\left( \frac{\partial(\sqrt{-\bar{g}}\overline{\mathcal{L}}_{c})}{\partial A_{\alpha}}-\nabla_{\mu} \frac{\partial(\sqrt{-\bar{g}}\overline{\mathcal{L}}_{c})}{\partial (\nabla_{\mu} A_{\alpha})}\right).\label{sec2:eqn11}
\end{equation}
This result is quite general in the sense that can be applied to more general vector-tensor theories, involving higher order derivatives self-interactions and non-minimal coupling to gravity, and can be extended to more general disformal transformations\footnote{For instance disformal transformations containing higher order derivatives of the vector field of the form $\bar{g}_{\mu \nu}=C(Y^{2},Y^{4})g_{\mu \nu}+B(Y^{2},Y^{4}) F_{\mu\rho}g^{\rho\sigma} F_{\sigma\nu}$ will contribute to terms beyond the second term of eqn.~(\ref{sec2:eqn11}) and, therefore, to a more general coupling. The equations of motions can be however reduced to second order by finding the associated Hamiltonian constraint \cite{Gumrukcuoglu:2019ebp}.}. The purpose of this paper however is to apply these results to a canonical vector-tensor theory consisting of the piece $\mathcal{L}_{2}(F_{\mu\nu},A_{\mu})$, in whose case such higher self-interactions are absent. Despite the minimal realization of this theory, it can exhibit interesting features in the dynamics of the universe due to the conformal and disformal couplings, as we shall see, since the interacting term depends essentially on the type of the transformation and not on the vector-tensor theory taken \textit{a priori}. On the other hand, in the present model the last term in equation eqn.~(\ref{sec2:eqn11}) vanishes since, by construction, there is no dependence of the barred metric on derivatives of the vector field (see eqn.~(\ref{sec2:eqn2})). This latter aspect is one of the most notorious difference in comparison to the vastly explored scalar disformal case. Thus, the chain rule allows us to rewrite the remaining part of eqn.~(\ref{sec2:eqn11}) in terms of the Jacobian transformation as follows
\begin{equation}
\frac{\partial}{\partial A_{\alpha}}\left(\sqrt{-\bar{g}} \overline{\mathcal{L}}_{\mathrm{c}}\right) =\frac{\partial\left(\sqrt{-\bar{g}} \overline{\mathcal{L}}_{\mathrm{c}}\right)}{\partial g_{\mu \nu}} \frac{\partial g_{\mu \nu}}{\partial \bar{g}_{\alpha \beta}} \frac{\partial \bar{g}_{\alpha \beta}}{\partial A_{\alpha}}=-\sqrt{-g} Q^{\alpha}.\label{sec2:eqn12}
\end{equation}
The Bianchi identities guarantee the covariant conservation of the total energy-momentum tensor
\begin{equation}
\nabla^{\mu} T_{\mu \nu}^{(A)}+\nabla^{\mu} T_{\mu \nu}^{(\mathrm{c})}=0,\label{sec2:eqn13}
\end{equation}
which, in turn, it is related to the Euler-Lagrange equation by virtue of the (first) Noether theorem
\begin{equation}
\begin{aligned}
\nabla_{\mu} T_{(A) \nu}^{\mu}=-\nabla_{\mu} T_{(c) \nu}^{\mu}&=&Q^{\mu} \nabla_{\nu} A_{\mu}-\nabla_{\mu}(Q^{\mu} A_{\nu})\\
&=&Q^{\mu}F_{\nu\mu}-A_{\nu}\nabla_{\mu}Q^{\mu}.\label{sec2:eqn14}
\end{aligned}
\end{equation}
It is also important to highlight that the source term in eqn.~(\ref{sec2:eqn14}) does not follow the same structure as its scalar analogue, i.e. it has not the compact form $Q\nabla_{\nu}\phi$ present in coupled scalar field models, but, on the contrary, it is more involved by virtue of the second term and because $Q^{\mu}$ is promoted to a tensor quantity. 

In a more fundamental physical ground, given that we expect that our low-energy world (late-time cosmology) is described by an effective field theory, as the low-energy description eqns.~(\ref{sec2:eqn1}) and (\ref{sec2:eqn2}), we do not expect that radiative corrections break the (effective) dark sector coupling at the scales we are concerned about. In other words, the theory we propose operates at sufficiently low-energy scale that quantum corrections are unimportant since the contributions from the non-renormalizable operators will be suppressed (far) below the strong coupling scale. Hence, the structure of the couplings remains untouched and the validity of the derived results are ensured at the associated low-energy scale.  Notice, however, that in the context of quintessence, a coupling to ordinary matter can rise even though it receives contributions from the theory at high-energies, which should lead to observable long-range forces \cite{Carroll:1998zi}.

\section{Concrete model}\label{sec:3}
We consider the standard Proca theory with a vector potential\footnote{The self-interacting potential plays mostly the same role as in the case of higher-order Lagrangians in the Generalized Proca theory (or other modified theories of gravity): provide self-accelerating solutions and, depending on the structure, contribute to the effective mass due to the presence of a massive vector field in gravity.}  for the vector-tensor sector $\mathcal{L}_{A}=m^{2}X+Y-V(X)$, so the coefficient containing higher-order derivative self-interactions $\mathcal{M}_{\beta}$ in eqn.~(\ref{sec2:eqn10}) vanishes. The artificial splitting of the potential and the mass term is done just to be reminiscent to the Generalized Proca theory where the mass term and the canonical Maxwell term belong to the lowest order Lagrangian ($\mathcal{L}_{2}=m^{2}X+Y$) and higher-order derivative self-interactions can be seen as corrections to the mass term \cite{Heisenberg:2014rta,Allys:2015sht}. Such derivative self-interactions for the vector field are precisely responsible of the existence of a self-accelerating solution \cite{DeFelice:2016yws} in a similar way the vector potential does in our case. Hence, a more general potential can include terms associated to derivative self-interactions that contribute to the effective mass due to the presence of a massive vector field in gravity.  Though this splitting is not necessary at all, it allows us to lie somehow in the spirit of modified gravity theories. We remind the results of section \ref{sec:2} are quite general and can be applied to more general vector-tensor theories. So, after explicit differentiation of $\mathcal{L}_{A}$, as indicated by the left-hand side of eqn.~(\ref{sec2:eqn10}), and calculating the interacting term according to eqn.~(\ref{sec2:eqn12}), the equation of motion for the vector field is reduced to the novel form
\begin{equation}
\begin{array}{l}
\nabla_{\mu} F^{\mu \nu}+(V_{,X}-m^{2}) A^{\nu}=-\frac{B}{C} 
 T_{(\mathrm{c})}^{\nu \mu} A_{\mu}+\frac{D}{2 C}(C-2 B X)\left(C_{,X} T_{(\mathrm{c})}+B_{,X} T_{(\mathrm{c})}^{\alpha \beta} A_{\alpha} A_{\beta}\right) A^{\nu},
\end{array}\label{sec3:eqn1}
\end{equation}
where we have defined the quantity $D\equiv\frac{1}{C-C_{,X}X+2B_{,X}X^{2}}$ in analogy to the scalar case. In the absence of coupling, that is $C=1$ and $B=0$, we recover the standard Proca theory plus a general potential. The energy momentum tensor of the vector field reads explicitly
\begin{equation}
\begin{array}{l}
 T^{\mu\nu}_{(A)}=F^{\mu}_{\sigma}F^{\sigma\nu}-\frac{1}{4}g^{\mu\nu}F^{\rho\sigma}F_{\rho\sigma}+m^{2}\left(A^{\mu}A^{\nu}-\frac{1}{2}g^{\mu\nu}A^{\rho}A_{\rho}\right)-V_{,X}A^{\mu}A^{\nu}-V g^{\mu\nu}.\label{sec3:eqn2}
 \end{array}
\end{equation}
%
We proceed now to compute the field equations in the FLRW spacetime with line element $ds^{2}=-dt^{2}+a^{2}(t)\delta_{ij}dx^{i}dx^{j}$, where $a(t)$ is the scale factor. To do so, we consider the commonly adopted temporal configuration for the vector field which is compatible with a homogeneous and isotropic background. Moreover, another reason why we choose such a particular configuration is because it can support the disformal (last) term in eqn.~(\ref{sec3:eqn1}), contrary to the purely spatial configuration (or cosmic triad), given that matter is assumed pressureless. Accordingly, we take
\begin{equation}
    A_{\mu}\equiv (A(t),0,0,0),\label{sec3:eqn3}
\end{equation}
to allow the generality of the coupling setting proposed. Here $A(t)$ is the temporal component of the vector field. Accordingly, the field equations read explicitly
\begin{equation}
3M_{p}^{2}H^{2}=\left(\frac{m^2}{2}-V_{X}\right)A^{2}+V+\rho_{c}+\rho_{r},\label{sec3:eqn4}
\end{equation}
\begin{equation}
M_{p}^{2}(3H^{2}+2\dot{H})=V-\frac{1}{2}m^{2}A^{2}-\frac{\rho_{r}}{3},\label{sec3:eqn5}
\end{equation}
\begin{equation}
(m^{2}-V_{X})A=\frac{A\rho_{c}(C_{X}-B_{X}A^{2}-2B)}{B_{X}A^{4}-C_{X}A^{2}+2C}.\label{sec3:eqn6}
\end{equation}
Here an upper dot denotes derivative with respect to cosmic time and $H(t)\equiv \dot{a}/a$ is the Hubble parameter. It is instructive to see that the branch $A=0$ in eqn~(\ref{sec3:eqn6}) is allowed as in the case of the standard Proca and Generalized Proca theories. Nevertheless, there can exist other solutions satisfying the equation of motion of the vector field in comparison to the uncoupled case\footnote{It implies that the coupling can enhance the vector field dynamics in periods when dark matter contribute significantly to the energy density of the Universe. It will be then quite interesting to investigate the model proposed in \cite{Tasinato:2014mia} where the equation of motion of the vector field is simply a constraint equation. Hence, once the coupling is turned on, the equation of motion can now evolve comprehensively to drive the cosmological acceleration beyond de Sitter solution found there.}. It means that in this simple scenario the coupling supports the time evolution of the vector field which depends clearly on the energy density associated to dark matter and the coupling functions. We expect then that the vector field vanishes during the radiation dominance or it does not play any role when dark matter is subdominant to the energy density of the universe. In other words, the coupling becomes ineffective and in turn the vector field, by construction, in regions of low dark matter density. From eqns.~(\ref{sec3:eqn4})-(\ref{sec3:eqn5}), we define the energy density $\rho_{A}$ and  pressure $P_{A}$ for the vector field
\begin{equation}
\rho_{A}=\left(\frac{m^2}{2}-V_{X}\right)A^{2}+V,\label{sec3:eqn7}
\end{equation}
\begin{equation}
p_{A}=\frac{m^2A^{2}}{2}-V.\label{sec3:eqn8}
\end{equation}
From these definitions we can derive the continuity equation associated to dark energy with equation of state $w_{A}=\frac{P_{A}}{\rho_{A}}$ and an interaction term\footnote{This term is exactly equal to the right hand side of eqn.~(\ref{sec2:eqn14}), so this is not the source term $Q$ in eqn.~(\ref{sec2:eqn10}). On the other hand, note that $\tilde{Q}$ contains time derivatives of $\rho_{c}$ but, after some algebraic manipulations, they all can be rewritten in terms of $\rho_{c}$ and rearranged to recover the canonical form of the continuity equation.} $\tilde{Q}$
\begin{equation}
\dot{\rho}_{A}+3H(\rho_{A}+p_{A})=-\tilde{Q}.\label{sec3:eqn9}
\end{equation}
Assuming a perfect-like fluid for dark matter in the Einstein frame we get 
\begin{equation}
\dot{\rho}_{c}+3H\rho_{c}=\tilde{Q}.\label{sec3:eqn10}
\end{equation}
It is convenient and possible to split the interaction term into the conformal and disformal contributions, for a better interpretation and treatment in the subsequent analysis, as $\tilde{Q}=\rho_{c} \frac{\dot{A}}{2A}\gamma=\rho_{c} \frac{\dot{A}}{2A}(\gamma_{C}+\gamma_{B})$, with
\begin{equation}
    \gamma_{C}=\frac{-2\frac{C_{X}}{C}A^{2}+A^{4}\left(\frac{C_{X}^{2}}{C^{2}}-2\frac{C_{XX}}{C}\right)}{\left(\frac{C_{X}}{C}A^{2}-2\right)\left(\frac{C_{X}}{C}A^{2}-1\right)},\quad \text{and}\quad \gamma_{B}=\frac{A^{2}\left(10B_{X}A^{2}+4B+A^{6}(B_{X}^{2}-2B_{XX}B)+2A^{4}(B_{XX}-3B_{X}B)\right)}{(2+B_{X}A^{4})(1+B_{X}A^{4}+BA^{2})}.\label{sec3:eqn11}
\end{equation}
Thus, the continuity equations tell us that both components interact with each other through a novel interaction term $\tilde{Q}$ determined by the purely conformal $\gamma_{C}$ and disformal $\gamma_{B}$ couplings. A very useful quantity that account for the evolution of the universe is the effective state parameter 
\begin{equation}
w_{\rm eff}\equiv \frac{p_{\rm T}}{\rho_{\rm T}}=-\left(1+\frac{2\dot{H}}{3H^{2}
}\right),\label{sec:3:eqn8}
\end{equation}
where $p_T$ and $\rho_T$ are respectively the total pressure and energy density. At this point we have derived all the key equations for the subsequent analysis of the background dynamics.

\subsection{Stability Analysis}\label{estability}
The coupling of the vector field to dark matter may in principle introduce some classical (and quantum) instability issues in the theory that can be associated to the presence of \textit{ghost fields}, that is, the propagation of undesired (physical) degrees of freedom. This subject has been indeed the central concern when building modified theories of gravity or theories with non-minimal couplings to matter that aim to go beyond general relativity.  There are some known analytical strategies to tackle this issue, such as the  Stueckelberg trick or perturbative analysis of linealized field equations around some  background spacetime. The former method sometimes allows a quick examination of instabilities of the scalar sector of the theory. We then adopt this approach. In doing so, we follow closely Ref.~\cite{Silva:2021jya}. that implemented the Stueckelberg trick to unveil ghost instabilities of vector fields in vectorized neutron stars\footnote{Several groups have also argued similar pathologies inherent to self-interacting vector fields \cite{Coates:2022qia,Coates:2023dmz,Clough:2022ygm,Mou:2022hqb} and, on the other side, possible solutions \cite{Aoki:2022woy,Barausse:2022rvg}.}.  
Let us first check the structure of the vector field equation of motion eqn.~(\ref{sec3:eqn1}). This can be recast however in a more canonical form. To do so, the generalized Lorentz constraint $g_{\nu\rho} \nabla^{\rho}(\hat{g}^{\mu\nu}A_{\mu})=0$ is derived by exploiting the antisymmetric property of $F_{\mu\nu}$  ($\nabla_{\nu}\nabla_{\mu}F_{\mu\nu}=0$), where the effective metric 
\begin{equation}
\hat{g}^{\mu\nu}=(V_{,X}^{\rm eff}+\beta) g^{\mu\nu} - \frac{B}{C}T_{c}^{\mu\nu},
\end{equation}
is defined for convenience. Here $V_{,X}^{\rm eff}=m^{2} - V_{,X}$ and $\beta=-\frac{D}{2 C}(C-2 B X)\left(C_{,X} T_{(\mathrm{c})}+B_{,X} T_{(\mathrm{c})}^{\alpha \beta} A_{\alpha} A_{\beta}\right)$.  Notice that  in the Lorenz constraint $\nabla_{\rho}\hat{g}^{\mu\nu}\neq0$.  At this point of the derivation the equation of motion can be rewritten in a compact way as 
\begin{equation}
\nabla_{\mu} F^{\mu \nu}=\hat{g^{\mu\nu}}A_{\mu},\label{eqn:EoM1}
\end{equation}
from which one can think mistakenly that $\hat{g^{\mu\nu}}$ corresponds to the effective mass squared, but it is not case, as we shall see below\footnote{This point was noticed in Ref.~\cite{Silva:2021jya} in vectorized solutions of neutron stars where tachyonic instability is invoked to trigger the existence of the vector field inside matter, leading inevitably to a pathological behavior. This aspect makes a clear distinction between both theories to advocate that our theory does not suffer from instability problems.}.  Interestingly,  in the more general case of disformal transformations the function accompanying $A_{\mu}$ in eqn.~(\ref{eqn:EoM1}) is promoted to a 2-rank tensor quantity and reduces to a scalar function as in the conformal case of Ref.~\cite{Silva:2021jya}. Finally, after rearranging cleverly eqn.~(\ref{eqn:EoM1}) to collect terms proportional to $A_{\mu}$ and expanding derivatives, one arrives at the wave equation\footnote{Notice that it is possible to eliminate derivatives of  $T_{(c)}^{\mu\nu}$ by using eqn.~(\ref{sec2:eqn14}).}
\begin{equation}
\Box A^{\rho} + \nabla_{\nu} \ln{\hat{z}} \;\nabla^{\rho}A^{\nu}-\nabla^{\rho} \left(\frac{B}{C}T_{(c)}^{\mu\nu} \nabla_{\nu}A_{\mu} \right) -  \nabla_{\nu}\left(\frac{B}{C}T_{(c)}^{\mu\nu}\right) \nabla^{\rho}A_{\mu}=\left[\nabla^{\rho}\left\{\nabla_{\nu} (\frac{B}{C} T_{(c)}^{\mu\nu})- \nabla^{\mu} \ln{\hat{z}}\right\} + R^{\rho\mu}+ \hat{g}^{\mu\rho} \right]A_{\mu},\label{eqn:EoM2}
\end{equation}
where we have defined $\hat{z}=(V_{,X}^{\rm eff}+\beta)$, and used the commutation rules for covariant derivatives. From here it is clear that the true effective mass squared corresponds to the quantity in the squared bracket.  This equation reduces consistently to eqn.~(6) of Ref.~\cite{Silva:2021jya} for the conformal case as can be checked.  One can consider only higher derivatives of the field  to see the principal part of the equation of motion which gives us, besides, a notion of its hyperbolic structure. Hence, first and third terms provide at leading order 
\begin{equation}
    \left(g^{\alpha\beta} g^{\mu\rho}-\frac{B}{C}T_{(c)}^{\mu\alpha}g^{\beta\rho}\right) \nabla_{\beta}\nabla_{\alpha}A_{\mu}+...= \mathcal{M}^{\mu\rho} A_{\mu},\label{eqn:leadingeqn} 
    \end{equation}
where $T_{(c)}^{\mu\alpha}=g^{\mu\alpha}diag(-\rho_{c},0,0,0)$ for a pressurless DM fluid, $...$ refers to low order derivatives 
of the vector field and $\mathcal{M}^{\mu\rho}$ accounts for the true effective tensor mass squared whose associated eigenvalues correspond to the masses of the physical degrees of freedom.  Hence, in order to keep the right sign of the kinetic energy and thus avoid the propagation of ghost modes, the coupling functions should have both the same sign during the whole cosmological evolution. On the other hand, gradient instabilities are trivially absent in a theory  where the vector field configuration is purely temporal (like eqn.~(\ref{sec3:eqn3})) and the spacetime background is isotropic and homogeneous\footnote{For instance, in a FRW background metric with a temporal vector field configuration that term has the specific form $(1+\frac{B}{C}\rho_{c}) \ddot{A}$.}. The signature of the coefficient in eqn.~(\ref{eqn:leadingeqn}) is also crucial to determine the hyperbolic character of the equation of motion which is ensured under the conditions mentioned above. In this sense, as long as the coefficient is well-behaved  (i.e. no singular) everywhere the field equation represents hyperbolic evolution. In short, in the pure conformal case ($B=0$) the theory is \textit{per se} free of ghost instabilities, and in the pure disformal case ($C=1$) the condition $B>0$ must be fulfilled. This is an important result we have to keep in mind in what follows.

The Stueckelberg field $\psi$ is usually introduced in a  theory to restore the gauge invariance and investigate the dynamical behavior of different degrees of freedom in the theory in question. Doing the substitution
\begin{equation}
A_{\alpha}\rightarrow A_{\alpha}+m_{V}^{-1}\nabla_{\alpha}\psi,\label{eqn:transf}
\end{equation}
the action is  recast in the form
\begin{equation}
\mathcal{S} =\int d^{4}x\left[\sqrt{-g} \left(\frac{M_{p}^2}{2}R -\frac{1}{4}F_{\mu\nu}F^{\mu\nu}-\frac{1}{2}g^{\mu\nu}(m_{V}A_{\mu}+\nabla_{\mu}\psi)(m_{V}A_{\mu}+\nabla_{\mu}\psi)+V(A_{\mu},\nabla_{\mu}\psi)\right) +\sqrt{-\bar{g}} \bar{\mathcal{L}}_{c}[\bar{g}_{\mu\nu} (A_{\mu},\nabla_{\mu}\psi),\psi_{c}]      \right].\label{eqn:stuekelbergaction}
\end{equation}
The Maxwell term is itself invariant under the transformation eqn.~(\ref{eqn:transf}). Variation of the new action with respect to the vector and scalar fields gives, respectively, the equations of motion
\begin{equation}
\nabla_{\mu}F^{\mu\nu}+(V_{,A}-m_{V})g^{\mu\nu}(m_{V}A_{\mu}+\nabla_{\mu}\psi)=-\frac{1}{2}\sqrt{C^{3}(C-2B X)}\Sigma^{\nu}\label{eqn:vectorfield}
\end{equation}
and
\begin{equation}
\Box\psi+(V_{,\psi}-m_{V})\nabla_{\mu} A^{\mu}=Q_{\psi},\label{eqn:scalarfield}
\end{equation}
where 
\begin{equation}
\Sigma^{\nu} = - \bar{T}_{(c)}^{\rho\sigma} (m_{V}^{2}A^{\nu}+m_{V}\nabla^{\nu}\psi) (g_{\rho\sigma}C_{,X}+B_{,X}  \chi_{\rho\sigma}m_{V}^{-2})+B (2 \bar{T}_{(c)}^{\sigma\nu} A_{\sigma}+m_{V}^{-1}\nabla_{\sigma}  \bar{T}_{(c)}^{\sigma\nu}),
\end{equation}
\begin{equation}
Q_{\psi}= \nabla_{\mu}\left[  \frac{1}{2} \sqrt{C^{3}(C-2BX)} \left\{  \frac{\partial \ln C}{\partial(\nabla_{\mu}\psi)} (\bar{T}_{(c)}-\frac{B}{m_{V}^{2}}\bar{T}^{\rho\sigma}\chi_{\rho\sigma}) + \frac{B}{m_{V}^{2}}  \left( \frac{\partial \ln B}{\partial(\nabla_{\mu}\psi)}\bar{T}^{\rho\sigma} \chi_{\rho\sigma} +2 \bar{T}^{\rho\mu}(m_{V}A_{\rho}+\nabla_{\rho}\psi)\right) \right\} \right],\label{eqn:interscalar}
\end{equation}
\begin{equation}
\chi_{\rho\sigma} =(m_{V}^{2}A_{\rho}+m_{V}\nabla_{\rho}\psi) (m_{V}^{2}A_{\sigma}+m_{V}\nabla_{\sigma}\psi).\label{eqn:scalarfactos}
\end{equation}
As can be seen the full theory is quite involved to be studied by  this method so we study the conformal case in what follows and leave for future work the disformal case. Hence, turning off the disformal part in eqn.~(\ref{eqn:scalarfield}), the scalar field equation becomes 
\begin{equation}
\Box\psi+(V_{,\psi}-m_{V})\nabla_{\mu} A^{\mu}=\nabla_{\mu}\left[ \frac{C^{2}}{2} \frac{\partial \ln C}{\partial(\nabla_{\mu}\psi)} \bar{T}_{(c)}  \right] = \nabla_{\mu}\left[ \beta  (m_{V} A^{\mu}+\nabla^{\mu}\psi) \right].\label{eqn:scalarfield2}
\end{equation}
Accordingly, the Lorenz constraint is reduced to the constraint equation $\nabla_{\nu}\left[ \tilde{z} (A^{\nu}+m_{V}^{-1}\nabla^{\nu}\psi)\right]=0$, where $\tilde{z}=1-V_{,X}/m_{V}^{2}+\beta/m_{V}^{2}$ and $\beta=-\frac{C_{,X}}{2C}\bar{T}_{(c)} C^{2}$. Hence if $\tilde{z}<0$, a tachyonic instability is developed (see eqn.~ (\ref{eqn:EoM1})).  This is simply avoided by taking, for instance, a general potential function of the form $V(X)=f(-\lambda X)$ with $\lambda>0$, and a general coupling function $C(X)=g(C_{0} X)$ with $C_{0}>0$ (or even $C_{0}<0$) such that $\beta>0$, since  $\bar{T}_{(c)}=-\bar{\rho}$.  Another more restrictive possibility is $|\beta|>|V_{,X}|$ but independent of the sign of $\lambda$. This condition is precisely the one we shall consider henceforth because it is consistent with the dynamical system constraints. Both conditions must be guaranteed however dynamically. Going further in the  analysis, let us focus on the scalar field equation. We introduce then an effective metric $\tilde{g}_{\mu\nu}=\tilde{z}^{-1}g_{\mu\nu}$ and rewrite the scalar field equation in terms of such a metric with the help of the Lorenz constraint from which we get the relation $\nabla_{\mu}(m_{V}A^{\mu}+\nabla^{\mu}\psi)=-(m_{V}A^{\mu}+\nabla^{\mu}\psi)\nabla_{\mu}\log\tilde{z}$. This yields
\begin{equation}
\tilde{\Box}\psi=-\tilde{g}^{\mu\nu}\left[(V_{,\psi}-m_{V})\nabla_{\mu} A_{\nu}+  V_{,XX}m_{V}^{-2} (m_{V}A_{\mu}+\nabla_{\mu}\psi) (m_{V}A_{\nu}+\nabla_{\nu}\psi) + (V_{,X} m_{V}^{-2}-1)\nabla_{\mu}\log{\tilde{z}} (m_{V}A_{\nu}+\nabla_{\nu}\psi) \right].\label{eqn:scalarfield3}
\end{equation}
Hence, in both representations, $g_{\mu\nu}$ and $\tilde{g}_{\mu\nu}$, the signature must keep fixed otherwise the field $\psi$ becomes a ghost at least in some region of the spacetime. This depends crucially whether $\tilde{z}$ changes sign. As we saw above, $\tilde{z}$ is always positive for the aforementioned conditions. Hence, the conformal part of theory is not prone to instabilities issues.

Thus, first examination tells us that if $B>0$ ghost instabilities are absent in the theory either in the pure disformal case or in the more general case. Even though we did not analyze the structure of the equation of motion of the scalar field in the more general case given by eqns.~(\ref{eqn:scalarfield}), (\ref{eqn:interscalar}) and (\ref{eqn:scalarfactos}), we speculate that the condition  $B>0$ is sufficient to avoid ghost instabilities at least in a FRW spacetime background with pure temporal configuration for the vector field. For instance, in the disformal case $\beta=-\frac{B_{X}A^{2}\rho_{c}(1-A^{2}B)}{2(1+\frac{B_{X}}{2}A^{4})}$, whereby  $B>0$ leaves the theory free of tachyonic instabilities ($\tilde{z}>0$) in a similar manner than the conformal case (see discussion just below eqn.~(\ref{eqn:scalarfield2})).
Finally, notice that instabilities can also occur dynamically whereby the whole evolution of the coupled system must be checked to determine the conditions under this could take place. This must be addressed numerically to avoid safely any pathological behavior of the theory.

\section{Dynamical system}\label{sec:4}
We proceed now to rewrite the system of eqns.~(\ref{sec3:eqn4})-(\ref{sec3:eqn6}) and eqns.~(\ref{sec3:eqn9})-(\ref{sec3:eqn10}) in the form of an autonomous system. It is convenient to define firstly the following dimensionless quantities\footnote{The introduction of $v$ is motivated mainly by two technical reasons: first, it allows us to trace and compact more easily terms proportional to the vector field which arise from the coupling of matter to the vector field and the vector potencial once their explicit forms are specified. Second, it helps us to close the system and write it in the form of an autonomous system. Hence, the choice of our variables renders the phase space compact without increasing the dimension.} that define the phase space portrait:
\begin{align}
& x \equiv \sqrt{\frac{-V_{X}A^{2}}{3M_{p}^{2}H^{2}}}; \;\; y\equiv \sqrt{\frac{V}{3M_{p}^{2}H^{2}}}; \;\; z \equiv \sqrt{\frac{\rho_{c}}{3M_{p}^{2}H^{2}}}; \;\; r \equiv \sqrt{\frac{\rho_{r}}{3M_{p}^{2}H^{2}}}; \;\; u\equiv \sqrt{\frac{m^{2}A^{2}/2}{3M_{p}^{2}H^{2}}}; \;\; v\equiv \frac{A}{M_{p}}.
\label{sec4:eqn1}
\end{align}
According to these definitions, the Friedmann constraint yields
\begin{equation}
x^{2}+y^{2}+z^{2}+r^{2}+u^{2}=1.\label{sec4:eqn2}
\end{equation}
With this, we are equipped to obtain the first order differential equations:
\begin{align}
x^\prime&=x\left(-\epsilon_{H}+\frac{v^\prime}{v}(1-\lambda v^{2})\right), \nonumber \\
y^\prime&=-y\left(\epsilon_{H} +\lambda v^\prime v\right),\nonumber \\
z^\prime&=\frac{z}{2}\left(-2\epsilon_{H}-3+\frac{\gamma}{2}\frac{v^\prime}{v}\right),\nonumber \\
r^\prime&=-r\left(2+\epsilon_{H}\right),\nonumber \\
\frac{u^\prime}{u}&= \frac{v^\prime}{v}-\frac{H^\prime}{H},\nonumber\\
\frac{v^\prime}{v}&=\frac{6u^{2}+3x^{2}}{4\lambda^{2}v^{4}y^{2}-x^{2}-2u^{2}-\frac{\gamma}{2}z^{2}}. \label{sec4:eqn3}
\end{align}
Here the prime denotes derivative with respect to $N\equiv\ln a$. It is interesting to point out that the last term in the differential equation of $z$ evidences the coupling between dark matter and the vector field which is also present in the equations associated to the vector field. Such a term enriches the overall dynamics of the cosmological model in comparison to the uncoupled case. In the above equations the accelerating equation
\begin{equation}
\epsilon_{H}=\frac{H^\prime}{H}=-\frac{3}{2}(1+w_{\rm eff})\quad \text{with}\quad w_{\rm eff}=\frac{r^{2}}{3}+u^{2}-y^{2},\label{sec4:eqn4}
\end{equation}
and an exponential potential $V(X)=V_{0}e^{-2\lambda X/M_{p}^{2}}$,
with $\lambda$ being a dimensionless model parameter, have been used. Note that from eqn.~(\ref{sec4:eqn1}) we can get the useful relation $x^{2}=2\lambda v^{2}y^{2}$ that allows us to reduce the dimension of the phase space since $v$ is a necessary variable to close the system. Finally, the equation of state parameter for dark energy is
\begin{equation}
w_{A}=\frac{-y^{2}+u^{2}}{x^{2}+y^{2}+u^{2}}.\label{sec4:eqn5}
\end{equation}
It remains to define the functional form of the coupling functions $C(X)$ and $B(X)$ entering in $\gamma$ through eqn.~(\ref{sec3:eqn11}). Hence, some particular forms for the couplings will be assumed in the next part to have concrete examples of how the emerging interaction operates at the level of the background. 
\begin{table*}[htp]
\centering  
\caption{Fixed points of the autonomous system described by eqn.~(\ref{sec4:eqn3}) for both type of conformal couplings chosen (eqn.~(\ref{sec4:eqn6})-(\ref{sec4:eqn6a})) and their main physical features such as the energy density parameter of the vector field (dark energy), its equation of state, the effective equation of state parameter, conditions for the existence of the critical points in phase space, and the conditions for supporting late-time accelerated expansion. Critical points marked with a tilde belong to the exponential coupling only but several critical points coexist in both cases. These are ($\rm A_{\pm}$), ($\rm B_{\pm}$), ($\rm E_{\pm}$), ($\rm D$) and ($\rm S$) solutions.
}
\begin{ruledtabular}
\begin{tabular}{ccccccccccc}
Point & $r_{c}$ & $y_{c}$ & $z_{c}$ & $u_{c}$ & $v_{c}$ & $\Omega_{A}$ & $w_{A}$ & $w_{\rm eff}$ & \text{Existence} & \text{Acceleration}
  \\ \hline
 $(\rm{A_{\pm}})$&$\pm 1$&$0$ &$0$&$0$ &$0$&  $0$ &$-$ &$1/3$ & $\forall q (\alpha),\lambda$ & $\text{No}$\\
 $(\rm{B_{\pm}})$&$0$
 &$0$&$\pm 1$&$0$ &$0$&  $0$ & $-$ & $0$ & $\forall q (\alpha),\lambda$ & $\text{No}$\\
 $(\rm{\tilde{B}_{1,2}})$&$0$
 &$0$&$\pm 1$&$0$ &$\mp\frac{1}{\sqrt{2\alpha}}$&  $0$ & $-$ & $0$ & $\alpha\neq0,\forall\lambda$ & $\text{No}$\\
 $(\rm{\tilde{B}_{3,4}})$&$0$
 &$0$&$\pm 1$&$0$ &$\pm\frac{1}{\sqrt{2\alpha}}$&  $0$ & $-$ & $0$ & $\alpha\neq0,\forall\lambda$ & $\text{No}$\\
 $(\rm{C_{1,2}})$&$0$&$0$
 &$\pm \sqrt{\frac{-2+6q}{-2+5q}}$&$\mp \sqrt{\frac{q}{2-5q}}$  &$0$& $\frac{q}{2-5q}$ & $1$ & $\frac{q}{2-5q}$ & $q\neq2/5, 0<q<1/3, \forall\lambda$ & $2/5<q<1$\\
  $(\rm{C_{3,4}})$&$0$&$0$
 &$\pm \sqrt{\frac{-2+6q}{-2+5q}}$&$\pm \sqrt{\frac{q}{2-5q}}$  &$0$& $\frac{q}{2-5q}$ & $1$ & $\frac{q}{2-5q}$ & $q\neq2/5, 0<q<1/3, \forall\lambda$ & $2/5<q<1$\\
 $(\rm{\tilde{D}_{1,2}})$&$0$&$1$
 &$\pm\sqrt{-\frac{\lambda}{\alpha}}$&$0$ &$\mp\frac{1}{\sqrt{2\alpha}}$& $1+\frac{\lambda}{\alpha}$ & $-\frac{1}{1 +\frac{\lambda}{\alpha}}$ & $-1$ & $\alpha\neq0, \lambda<0$ & $\text{Yes}$\\
  $(\rm{\tilde{D}_{3,4}})$&$0$&$1$
 &$\pm\sqrt{-\frac{\lambda}{\alpha}}$&$0$ &$\pm\frac{1}{\sqrt{2\alpha}}$& $1+\frac{\lambda}{\alpha}$ & $-\frac{1}{1 +\frac{\lambda}{\alpha}}$ & $-1$ & $\alpha\neq0, \lambda<0$ & $\text{Yes}$\\
  $(\rm{\tilde{F}_{1,2}})$&$0$&$1$
 &$\pm\sqrt{-\frac{\lambda}{2\alpha}}$&$0$ &$\mp\frac{1}{2\sqrt{\alpha}}$& $1+\frac{\lambda}{2\alpha}$ & $-\frac{1}{1 +\frac{\lambda}{2\alpha}}$ & $-1$ & $\alpha\neq0, \lambda<0$ & $\text{Yes}$\\
  $(\rm{\tilde{F}_{3,4}})$&$0$&$1$
 &$\pm\sqrt{-\frac{\lambda}{2\alpha}}$&$0$ &$\pm\frac{1}{2\sqrt{\alpha}}$& $1+\frac{\lambda}{2\alpha}$ & $-\frac{1}{1 +\frac{\lambda}{2\alpha}}$ & $-1$ & $\alpha\neq0, \lambda<0$ & $\text{Yes}$\\
 $(\rm{E_{\pm}})$&$0$&$0$
 &$0$&$\pm1$ &$0$&  $1$ & $1$ & $1$ & $\forall q (\alpha),\lambda$ & $\text{No}$\\
$(\rm{D_{1,2}})$ &$0$&$\pm\frac{1}{\sqrt{1+v^{2}\lambda}}$
 &$0$&$\mp v\sqrt{\frac{\lambda}{-1-v^{2}\lambda}}$ &$-$& $1$ & $-1$ & $-1$ & $\forall q (\alpha), \text{eqn.~(\ref{sec4:eqn8})}$ & $\forall q (\alpha), \lambda$\\
 $(\rm{D_{3,4}})$ &$0$&$\pm\frac{1}{\sqrt{1+v^{2}\lambda}}$
 &$0$&$\pm v\sqrt{\frac{\lambda}{-1-v^{2}\lambda}}$ &$-$& $1$ & $-1$ & $-1$ & $\forall q (\alpha), \text{eqn.~(\ref{sec4:eqn8})}$ & $\forall q (\alpha), \lambda$\\
 $(\rm{S})$ &$0$&$1$
 &$0$&$0$ &$0$ & $1$ & $-1$ & $-1$ & $\forall q (\alpha), \lambda$ & $\forall q (\alpha), \lambda$\\
\end{tabular}
\end{ruledtabular}\label{table1}
\end{table*}
%

\subsection{Critical points and stability}\label{sec:4.1}
We follow the standard procedure to calculate the critical points\footnote{We strongly suspect that the existence of some unphysical solutions in this model is due to the non-trivial coupling of the vector field to matter. The existence of such solutions arise commonly in theories with non-canonical fields, non-minimal coupling to gravity and non-trivial couplings between different fields.}, that is, by matching to zero each equation of the autonomous system and solving a set of simple algebraic expressions. However, before doing so, it is necessary to define the conformal and disformal couplings in the autonomous system as discussed.
However, no matter their functional forms,  they are still free of the ghost-like Ostrogradski instability and can not be determined, as far as we know, by some physical principle beyond the assumptions made here\footnote{This kind of assumption has to be made in theories with free functions, as in the Generalized Proca theory where the free functions $G_{i}$ are taken ad-hod to realize the de Sitter fixed point while they are consistent with the stability conditions \cite{DeFelice:2016yws}.}.

Once some functional forms are given, the stability of the critical points will be analyzed separately in order to track the effects of each type of coupling in the dynamical behavior of the system as identified in eqn.~(\ref{sec3:eqn11}). 

\subsubsection{Conformal Case}\label{sec:4.1.1}

This case corresponds to $B(X)=0$, leading to $\gamma_{B}=0$ everywhere. We first assume a power law function for the conformal coupling
\begin{equation}
C(X)=C_{0}\left(\frac{X}{M_{p}^{2}}\right)^{q},  \label{sec4:eqn6}
\end{equation}
where $C_{0}$ and $q$ are constants. We will refer to this particular choice for the coupling function as \textit{conformally coupled power-law model} henceforth. For this model, the interaction function in the continuity equations takes the simple constant form $\gamma_{C}=\frac{2q}{1-2q}$. Another possibility we will explore is the exponential coupling
\begin{equation}
C(X)=\tilde{C}_{0}\;e^{\frac{4\alpha X}{M_{p}^{2}}},  \label{sec4:eqn6a}
\end{equation}
where $\tilde{C}_{0}$ and $\alpha$ are constants. We call this model 
\textit{conformally coupled exponential model} henceforth. The interaction function turns out to be now a function of the vector field
\begin{equation}
    \gamma_{C}=-\frac{4\alpha v^{2}(1+2v^{2}\alpha)}{1-6\alpha v^{2}+8\alpha^{2} v^{4}}.
\end{equation} 
With all this, the system is completely determined, it means that the physical space renders compact and close. We report then all the critical points in Table~\ref{table1} and show the conditions that determine both their dynamical character and their existence in phase space. Also, some physical quantities of interest are shown for a better comprehension of the dynamical behavior of the system. Both models can be analyzed in most of the cases (but carefully) together since they share some similarities in phase space as the existence of the fixed points $\rm (A_{\pm}), (B_{\pm}), (E_{\pm}), (D_{1,2}), (D_{3,4})$ and $\rm (S)$ for both types of couplings. Some of them also share the same eigenvalues except for $\rm (E)$. For these points the conditions for existence and acceleration associated to the exponential coupling, parameterized by $\alpha$, are shown in parentheses. The other points belong either for one coupling or another as can be read off. In particular, points marked with a tilde correspond to the exponential coupling only. Having specified which of the fixed points come from one or another coupling function, we discuss the main physical features of each of the fixed points as follows\footnote{Note that at late times there is not distinction between the vector field and dark energy nominations, so we will speak indistinctly when referring to them.}:
\begin{itemize}
\item \textbf{Point} ($\rm A_{\pm}$): this point describes the standard radiation dominance with a saddle-like behavior. Eigenvalues are independent of the model parameters $(-3, -1, 1/2, 2)$. On the other hand, it is not surprising the nonexistence of a scaling vector radiation solution, contrary to its scalar analogue, because the kinetic term $Y$ vanishes for a purely temporal configuration in FRLW background, which does not occur, on the contrary, for purely spatial configuration of the vector field. Put it in another way, the vector degree of freedom is not propagating.

\item \textbf{Point} ($\rm B_{\pm}$): this fixed point account for fully matter domination and is a saddle point with eigenvalues $(3/2, 3/2, -1, 0)$ and, as in the radiation case, it does not depend on the models parameters by any means.

\item \textbf{Point} ($\rm \tilde{B}$): despite the vector field does not vanish in this solution, physically things are not much different in this solution in comparison to the standard dark matter dominated scenario ($\rm B_{\pm}$), they even have the same eigenvalues.

\item \textbf{Point} ($\rm C$): this fixed point has the form of a scaling solution because of the absence of the potential parameter $\lambda$ that can generate acceleration and, even more, because of the presence of dark energy during the dark matter domination epoch. Nevertheless, the dynamical character of this fixed point must be established by checking the sign of the real parts of the eigenvalues of the Jacobian matrix associated to the linear system. Eigenvalues can be reduced to the form
\begin{equation}
    \left(\frac{3 - 6 q}{-2 + 5 q}, \frac{5-17q}{2(-2+5q)}-\chi,  \frac{5-17q}{2(-2+5q)}+\chi, \frac{3(-1+2q)(1+v^{2}\lambda)}{-2+5q}\right),\label{sec4:eqn7}
\end{equation}
where $\chi=\sqrt{\frac{-(2-5q)(-1+q)^{2}q(-2+5q)^{8}}{4q(-2+5q)^{11}}}$. There exist conditions in which eigenvalues have all negative real part, guaranteeing thereby the stability of this solution, this is, the attractor-like character of this point, while they still being consistent with both the conditions for acceleration ($2/5<q<1$) and  existence of the critical points ($0<q<2/5$). This point would correspond to an \textit{attractor solution} supported by the power coupling function. Nevertheless, as we can infer from the above ranges, their parameter spaces are, unfortunately, incompatible with each other. So this point is discarded to address the late-time accelerated expansion. On the other hand, it is easy to see that all eigenvalues can be either negative or positive, depending on the multiple choices of the model parameters. It can not be however a \textit{repeller} (eigenvalues with positive real parts) because the condition for the first eigenvalue to be positive ($2/5<q<1/2$) is not compatible with any of the other eigenvalues: $1/4<q<2/5$ for the second eigenvalue and $1/4<q<2/5$ for the third one. As this fixed point is already discarded as an attractor point and a repeller, we focus instead on the possibility of having a saddle point by demanding that at least one of the eigenvalues has opposite sign. After exploring the available parameter space we see that second and third eigenvalues are monotonically increasing functions for the range $0<q<1/3$ which cross the zero before $2/5$, and the first eigenvalue is a  monotonically decreasing function for the same range. So the latter has opposite sign at the time the other eigenvalues cross the zero. No matter actually whether this happens or not because the fourth eigenvalue has always opposite sign (positive) with respect to the first negative eigenvalue. This is also true for reasonable values of $v$ and $\lambda$ in eqn.~(\ref{sec4:eqn7}) as we have checked. Hence, this fixed point can be classified as \textit{saddle point}. Also, notice that the branch $\lambda=0$ (which leaves $v$ unconstrained) is allowed. It means that the exponential potential parameter may or may not affect the dynamical behavior of this fixed point. We can then conclude that this fixed point corresponds to a novel \textit{vector-dark matter scaling solution} with effective equation of state parameter $w_{\rm eff}=\frac{q}{2-5q}$ which deviates from zero for $q\neq0$ and increases monotonically with it. A small value of $q$ is then expected so that $w_{\rm eff}$ is close to zero (dark matter domination). This aspect will be better analyzed in the numerical analysis of the model taking,  \textit{a priori}, the inferred range $0<q<1/3$. Notice also that one can go directly to the standard dark matter domination point ($\rm B_{\pm}$) by making $q=0$, however this branch does not give rise to a new solution because it is not supported by a constant coupling; $\gamma_{C}$ is powered, instead, by derivatives of the coupling function (see eqn.~(\ref{sec3:eqn11})). On the other hand, the energy densities associated to dark matter and dark energy are, respectively, $\Omega_{\rm DE}=\frac{q}{2-5q}$ and $\Omega_{\rm DM}=\frac{-2+6q}{-2+5q}$. Here the mass term of the vector field is the one that supports the presence of dark energy during dark matter domination.

\item \textbf{Point} ($\rm S$): This fixed point is also present in both type of coupling but independent on the respective coupling parameters. This solution is a de Sitter attractor point with negative eigenvalues $(-3, -4, -3, -3/2)$. As this fixed point is fully supported by the exponential potential, it will be interesting to see deviations from this solution, as those provided by ($\rm D$) and ($\rm\tilde{D}$), though the latter seems to deviate more considerably from a constant dark energy density as we will see.

\item \textbf{Points} ($\rm \tilde{D}$, $\rm \tilde{F}$): these fixed points are scaling solutions modulated by the exponential coupling parameter $\alpha$ and both may account, in principle, for the accelerated expansion ($w_{\rm eff}=-1$). Despite subtle differences in their critical points (a factor of 1/2), their physical parameters differ roughly by a factor of a half so they can be analyzed together. The vector field contributes to the dark energy density so that $\Omega_{\rm DE}=1-\frac{|\lambda|}{2\alpha}$ and, therefore, $\Omega_{\rm DM}=\frac{|\lambda|}{2\alpha}$ in accordance with the conditions for the existence. For $\lambda=0$ we recover the de Sitter point $(\rm S)$ described above. Stability must be treated however separately. The fixed point ($\rm \tilde{F}$) is actually a \textit{saddle point} with eigenvalues $(0, -4, -3, 4)$ and the fixed point ($\rm \tilde{D}$) does have an \textit{attractor-like character} since all nonzero eigenvalues are negative, taking the simple form $(0,-4,-3,-3)$, and they all being independent of the model parameters. Stability is then guaranteed trivially without imposing further conditions\footnote{Note however that linear stability analysis fails to determine the stability properties of non-hyperbolic points whereby other alternative approaches must be implemented (see e.g. \cite{Bahamonde:2017ize}). We have used a heuristic criterion to confirm the attractor character of this kind of points by assessing whether different trajectories for a wide range of initial conditions in phase space converge ultimately to the conjectured attractor point. Although this is not shown here,  it can be checked analogously with the numerical analysis we present later.}.

\item \textbf{Point} ($\rm D$): this fixed point also provides accelerated expansion ($w_{\rm eff}=-1$) and is present in both type of couplings. Critical points depend on the exponential potential parameter $\lambda$ and the vector field $v$. This latter can take in principle any value to satisfy simultaneously the Friedmann constraint and the autonomous system but can be constrained  tightly from the condition for the existence of the critical points itself: 
\begin{equation}
    \lambda<0\; \land \; \left( -\sqrt{-\frac{1}{\lambda}}<v<0\; \lor \; 0<v<\sqrt{-\frac{1}{\lambda}}\right).\label{sec4:eqn8}
\end{equation}
Turning off the vector field $v=0$, leads to de Sitter point $(\rm S)$. Despite the eigenvalues are very lengthy, these can be reduced, after some manipulations and after exploring the suitable parameter space, to the simple form $(-\frac{3}{2},-3,-4,0)$ without loss of generality. Stability is also guaranteed without problems.

\item \textbf{Point} ($\rm E_{\pm}$): This fixed point is present regularly in many cosmological models based on scalar and vector fields and is supported by their kinetic energies. From here it is called kinetic dominated solution or kination in short. This is a saddle point with eigenvalues $(-3, 3, 2, 3/2)$ and $(-3, 2, \frac{3 - 9 q}{2 - 4 q}, 3)$, respectively, for the exponential and power law couplings. For our case, however, it is supported entirely by the mass term $u_{c}=1$ (since $Y=0$) and corresponds to a fully vector field domination with a stiff equation of state $w_{\rm eff}=1$, similar to the usual kinetic dominated solution. 

\end{itemize}

\begin{table*}[htp]
\centering
\caption{Fixed points of the autonomous system given by eqn.~(\ref{sec4:eqn3}) for the disformal coupling case with $\beta=-1/2$ and $\beta=-2$ choices. Fixed points marked with tilde belong to the  $\beta=-2$ sub case. Their main physical features such as energy density parameter of the vector field (dark energy), its equation of state, the effective equation of state parameter, conditions for the existence of the critical points in phase space, and the conditions for supporting late-time accelerated expansion are showed as well.}
\begin{ruledtabular}
\begin{tabular}{ccccccccccc}
Point & $r_{c}$ & $y_{c}$ & $z_{c}$ & $u_{c}$ & $v_{c}$ & $\Omega_{A}$ & $w_{A}$ & $w_{\rm eff}$ & \text{Existence} & \text{Acceleration}
  \\ \hline
  $(\rm{H_{1,2}})$&$0$&$\pm1$ &$\pm2\frac{\sqrt{-2\lambda}}{B_{0}}$&$0$ &$\mp\frac{2}{B_{0}}$&  $1+\frac{8\lambda}{B_{0}^{2}}$ &$-\frac{1}{1 +\frac{8\lambda}{B_{0}^{2}}}$ &$-1$ & $B_{0}\neq0, \lambda<0$ & $\text{Yes}$\\
 $(\rm{H_{3,4}})$&$0$&$\pm1$ &$\pm2\frac{\sqrt{-2\lambda}}{B_{0}}$&$0$ &$\mp\frac{2}{B_{0}}$&  $1+\frac{8\lambda}{B_{0}^{2}}$ &$-\frac{1}{1 +\frac{8\lambda}{B_{0}^{2}}}$ &$-1$ & $B_{0}\neq0, \lambda<0$ & $\text{Yes}$\\
 $(\rm{G_{1,2}})$&$0$
 &$0$&$\pm 1$&$0$ &$\pm\frac{2}{B_{0}}$&  $0$ & $-$ & $0$ & $B_{0}\neq0$ & $\text{No}$\\
 $(\rm{G_{3,4}})$&$0$
 &$0$&$\pm 1$&$0$ &$\mp\frac{2}{B_{0}}$&  $0$ & $-$ & $0$ & $B_{0}\neq0$ & $\text{No}$\\
 $(\rm{\tilde{H}_{1,2}})$&$0$&$\pm1$ &$\pm\sqrt{-2c B_{0}\lambda}$&$0$ &$\mp\sqrt{cB_{0}}$&  $1+2cB_{0}\lambda$ &$-\frac{1}{1 + 2cB_{0}\lambda }$ &$-1$ & $B_{0}>0, \lambda<0$ & $\text{Yes}$\\
 $(\rm{\tilde{H}_{3,4}})$&$0$&$\pm1$ &$\pm\sqrt{-2c B_{0}\lambda}$&$0$ &$\pm\sqrt{cB_{0}}$&  $1+2cB_{0}\lambda$ &$-\frac{1}{1 + 2cB_{0}\lambda }$ &$-1$ & $B_{0}>0, \lambda<0$ & $\text{Yes}$\\
 $(\rm{\tilde{G}_{1,2}})$&$0$
 &$0$&$\pm 1$&$0$ &$\pm\sqrt{cB_{0}}$&  $0$ & $-$ & $0$ & $B_{0}>0$ & $\text{No}$\\
 $(\rm{\tilde{G}_{3,4}})$&$0$
 &$0$&$\pm 1$&$0$ &$\mp\sqrt{cB_{0}}$&  $0$ & $-$ & $0$ & $B_{0}>0$ & $\text{No}$\\
\end{tabular}
\end{ruledtabular}\label{table2}
\end{table*}

\subsubsection{Disformal Case}\label{sec:4.1.2}

\begin{figure*}
\centering
\includegraphics[width=0.45\hsize,clip]{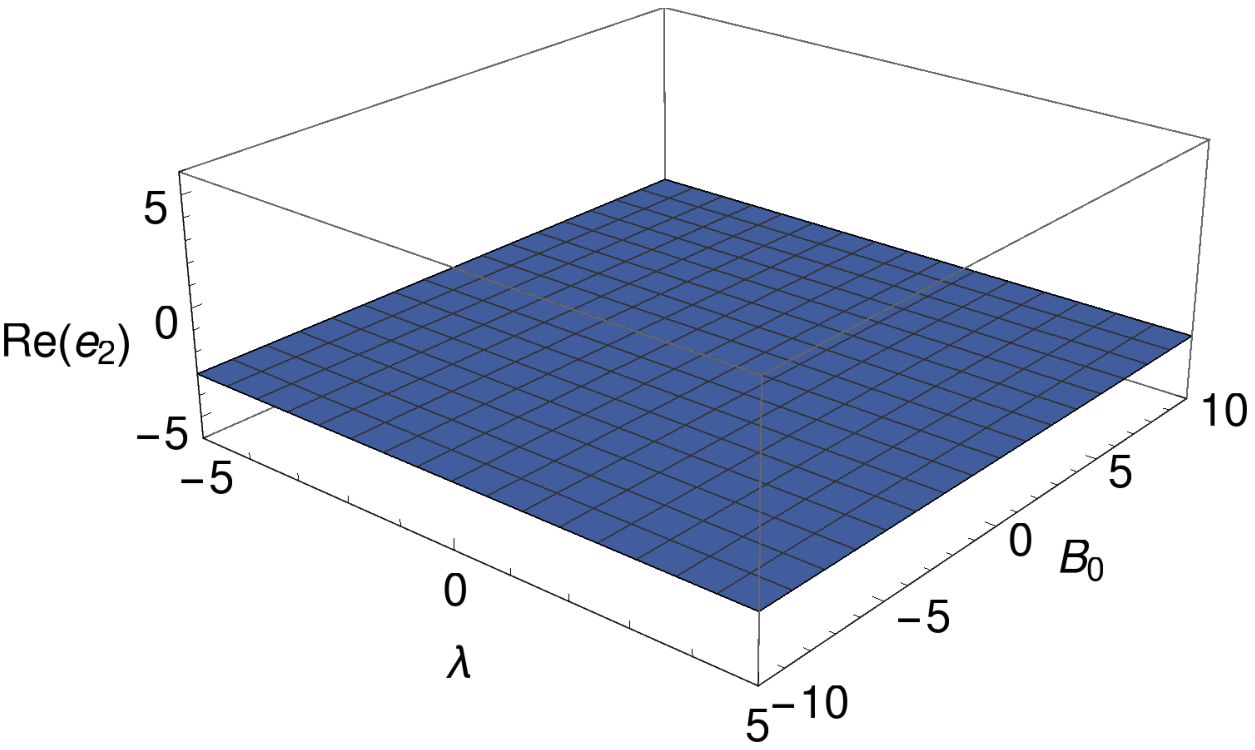}
\includegraphics[width=0.45\hsize,clip]{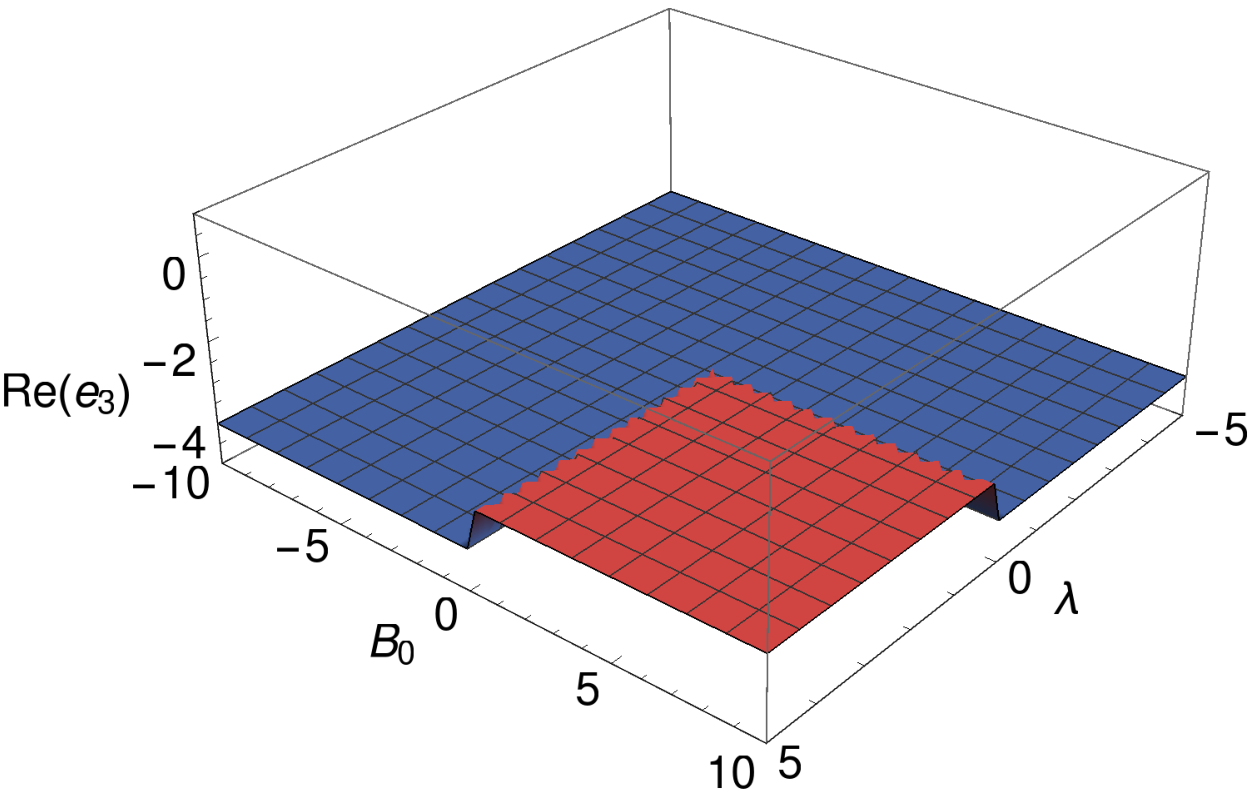}
\includegraphics[width=0.45\hsize,clip]{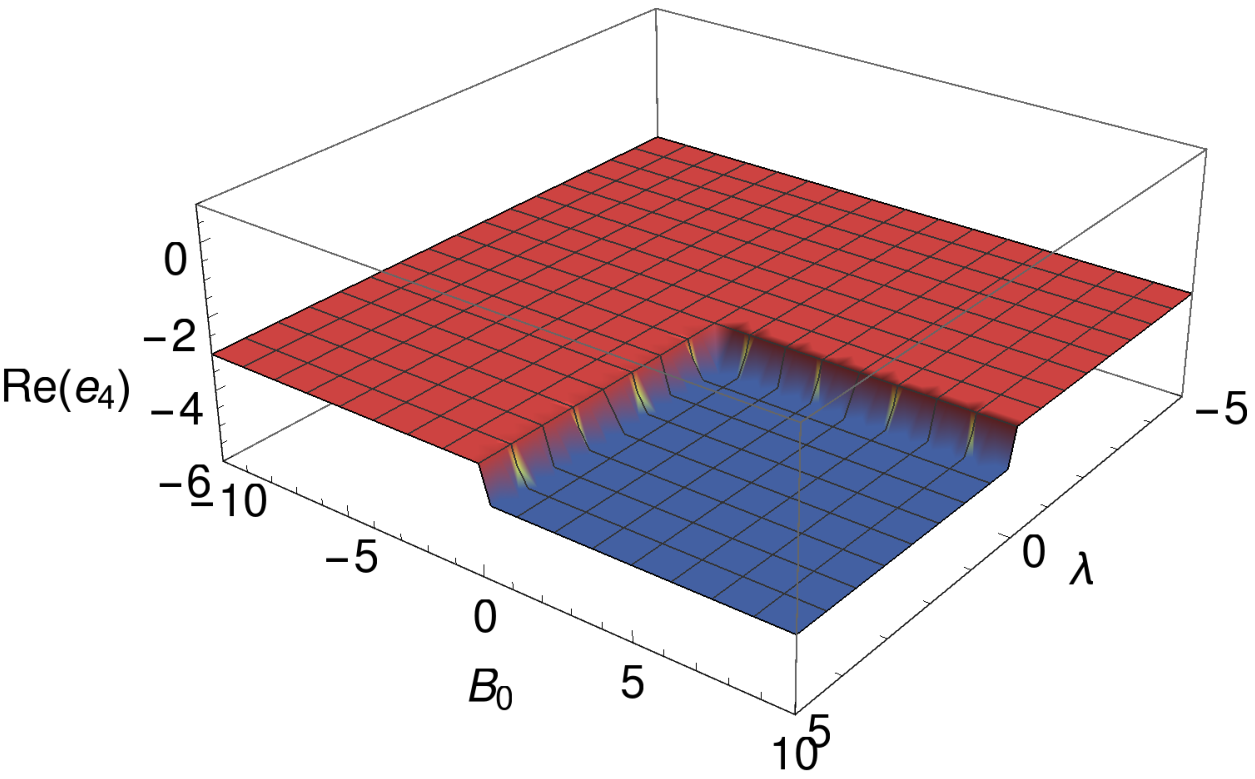}
\caption{Real part of eigenvalues associated to the fixed point $\rm (H)$ in the parameter space. These are always negative, so this solution is stable, corresponding to an attractor solution.} \label{fig:1}
\end{figure*}

\begin{figure*}
\centering
\includegraphics[width=0.45\hsize,clip]{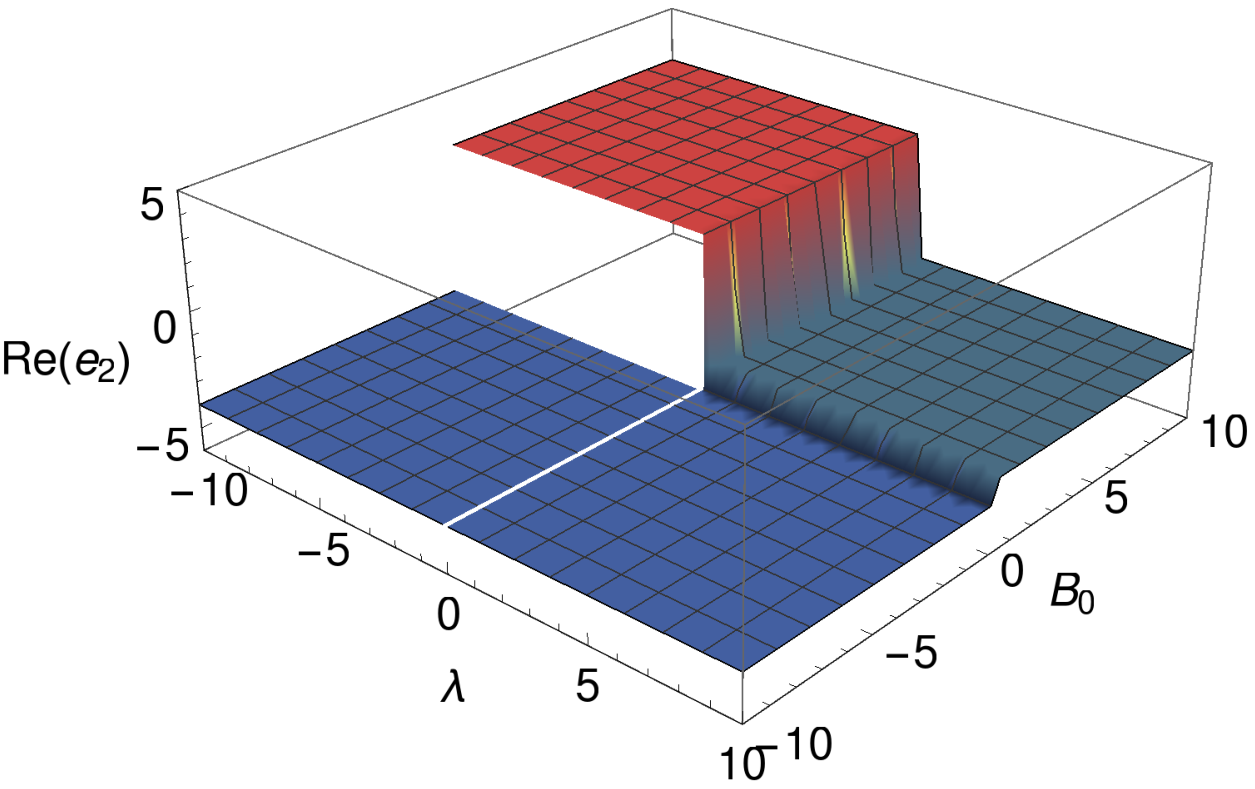}
\includegraphics[width=0.45\hsize,clip]{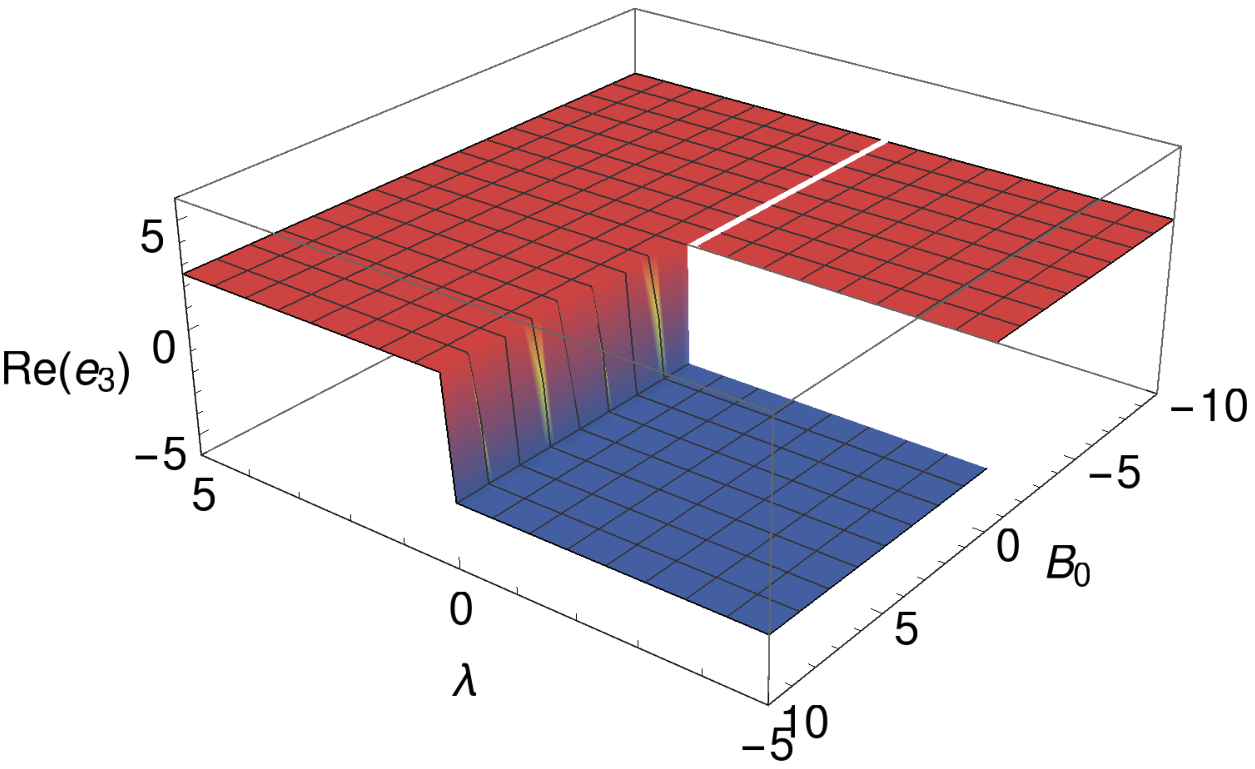}
\includegraphics[width=0.45\hsize,clip]{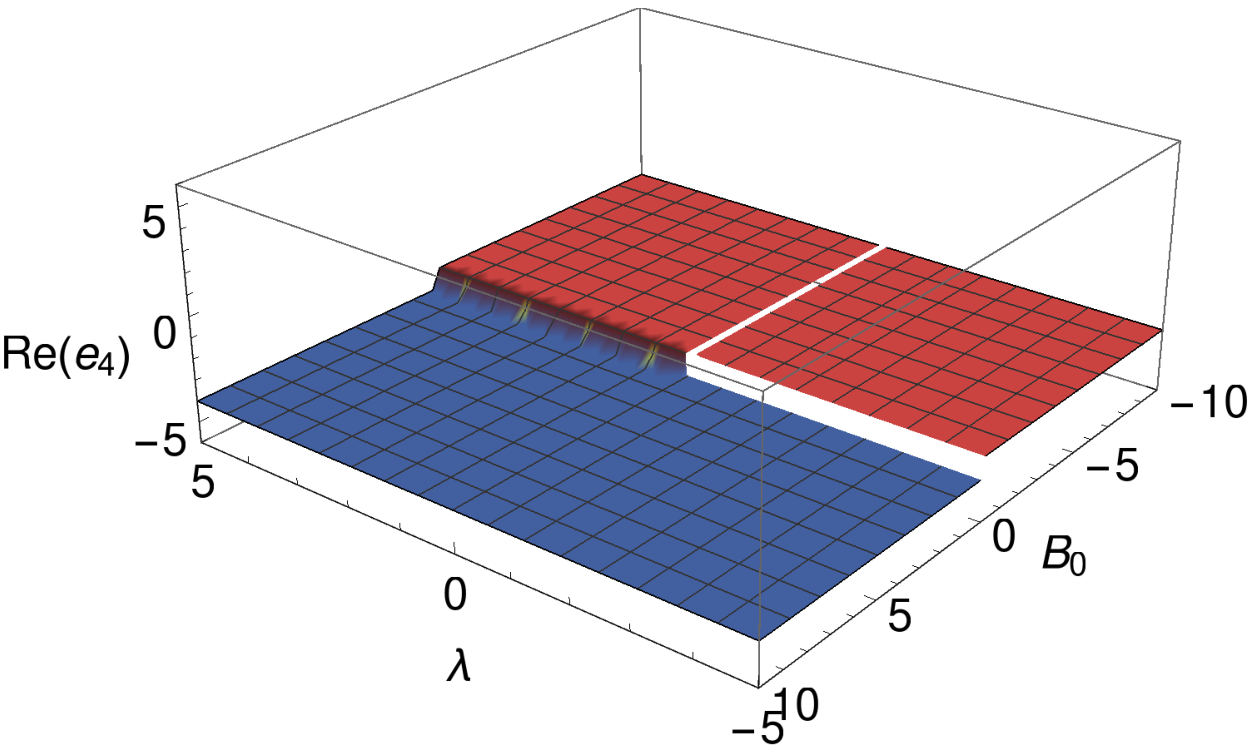}
\caption{Real part of eigenvalues associated to the fixed point $\rm\tilde{(H)}$ in the parameter space. These can not be simultaneously negative or positive for the same parameter values, so this solution corresponds to a saddle point.}\label{fig:2}
\end{figure*}

We can proceed in different ways, but looking for the effects of the pure disformal coupling and keeping a reasonable number of free parameters. Thus, we take for this model
\begin{equation}
    C(X)=1,\;B(X)=B_{0}\frac{2^{\beta}X^{\beta}}{M_{p}^{2+2\beta}},\label{sec:4:eqn0}
\end{equation}
where, unlike the conformal coupling, the disformal function has units of inverse energy squared. For this particular disformal function, the interaction term takes the non illuminating form
\begin{equation}
    \gamma_{B}=-B_{0}\frac{2 v^{2+2\beta}(1+\beta) \left(-1+\beta(-2+B_{0}v^{2+2\beta})\right)}{(1+\beta B_{0}v^{2+2\beta})(1+v^{2+2\beta}(1+2\beta)B_{0})},\label{sec:4:eqn1}
\end{equation}
which is clearly more involved compared to the conformal case. So far we have attempted to keep the generality in our analysis but, unfortunately, keeping $\beta$ free there exist many critical points that make this analysis intractable from the analytical point of view. We comment some possible choices after having explored the suitable parameter space for cosmological implications. For the coupling constant case $\beta=0$ there are no new critical points compared to the uncoupled case, though it does not mean that they can not affect the background dynamics as we will see later in the numerical analysis. $\beta=-1$ leads to the uncoupled case $\gamma_{B}=0$ whose critical points were analyzed altogether with the conformal case. That case can also be achieved by doing, of course, $B_{0}=0$. For $\beta$ positive most of the solutions are complex and therefore discarded. So, we focus mainly on $\beta$ negative with $\beta=-1/2,-2$. The reason why we take this particular values is because they encompass, after thorough examination, most of the physical solutions of interest within the available parameter space. Thus, $\beta=-1/2,-2$ provide, respectively, $\gamma_{B}=\frac{B_{0}^{2}v^2}{2-B_{0}v}$ and $\gamma_{B}=\frac{2B_{0}(3v^2-2B_{0})}{v^{4}-5v^{2}B_{0}+6B_{0}^{2}}$. Solving the autonomous system for these interaction terms give rise to several sets of critical points within which some of them have been already discussed in the uncoupled and conformal cases. These cover the standard radiation and matter dominated solutions ($A_{\pm}$) and ($\rm B_{\pm}$), respectively, the fixed point ($\rm D$) that supports accelerated expansion, the de Sitter solution ($\rm S$) and the fixed point ($\rm E$). So we analyze new emerging solutions characterized by the disformal coupling only. They are shown in Table~\ref{table2} as well as their main cosmological features. For the case $\beta=-2$ many more critical points appear in phase space in comparison to the case $\beta=-1/2$. The solutions are written in a compact way in terms of the parameter $c$, as defined below, for the sake of simplicity. In the following,  we discuss the dynamical character and the criteria for stability conditions.
\begin{itemize}

\item \textbf{Point} ($\rm H$): this fixed point corresponds to a \textit{stable attractor solution} with $\beta=-1/2$ and eigenvalues $(0,-4, -3, -3)$, they 
all having (non-zero) negative real parts, so stability is ensured straightforwardly. We think of this case as the minimal realization of the disformal model given the simple form of both critical points and eigenvalues compared to other values of $\beta$ that lead to a more involved stability conditions. This solution can then drive the late-time accelerated expansion with $w_{\rm eff}=-1$. From the condition of existence of the critical points we get $\lambda<0$ which allows us to write the energy density parameters as $\Omega_{\rm DM}=\frac{8|\lambda|}{B_{0}^{2}}$ and $\Omega_{\rm DE}=1-\frac{8|\lambda|}{B_{0}^{2}}$. This functional form is reminiscent of the energy density parameters associated to the fixed point $(\rm \tilde{D})$: $B_{0}^2$ is exchanged by $16\alpha$. The equation of state for dark energy reads $w_{\rm DE}=-\frac{1}{1+\frac{8\lambda}{B_{0}^{2}}}$ which depends on both parameters. For $\lambda$ going to zero, this fixed point tends to de Sitter solution; in this sense, therefore, $\lambda$ supports the coupling and their effects on the dynamic evolution from the dynamical system perspective. 

\item \textbf{Point} ($\tilde{\rm H}$): this solution, with $\beta=-2$, actually corresponds to two distinct physical solutions but they can be written in a compact way in terms of the parameter $c$ (with $c=2,3$), since their main physical properties can be analyzed together though it loses its validity when studying the stability conditions. Hence,  the dynamical character of this point must be treated separately to determine which of the aforementioned values of c correspond, or not, to a stable solution. This requirement is set by demanding that their associate eigenvalues have all negative real parts. As they are very lengthy to be reported here and be treated analytically, since they exhibit explicit dependence on the two parameters $B_{0}$ and $\lambda$ in non-trivial way, we adopt another strategy to establish the stability. Before moving on, the conditions $B_{0}\neq0$ and $\lambda \neq0$ must be guaranteed everywhere to allow eigenvalues take well-defined values. We notice, when plotting the real part of the eigenvalues, that they all form a series of constant planes whose values depend on the region of the parameter space whereby they can be recast, for the entire parameter space, in a parameterized way as follows.

For the case $c=2$, eigenvalues can be written as $(e_{1}, e_{2}, e_{3}, e_{4})$ with $e_{1}=0$. From figure (\ref{fig:1}), we infer the value $e_{2}=-3$ for any value of the model parameters $\lambda$ and $B_{0}$, $e_{3}=-3$ for $B_{0}>0$ and $\lambda>0$, or $e_{3}=-4$ for all other cases, among which the one consistent with the condition for the existence of the critical points. Likewise, $e_{4}=-4$ for $\lambda>0$ and $B_{0}>0$, or $e_{4}=-3$ for other cases. Here is also included the condition for existence. These eigenvalues are always negative anyway\footnote{We have also evaluated the numerical value of all eigenvalues in the discussed ranges of the parameter space to be sure that they correspond effectively to the ones we inferred from the plots, finding thus consistency between both approaches.}. Accordingly, any trajectory in phase space leaving the matter dominated period characterized by a saddle-type behavior will end up in this stable fixed point. Hence, this fixed point corresponds to an \textit{stable attractor solution} and is refereed to as \textit{disformally coupled dark energy solution} with $w_{\rm eff}=-1$, i.e., this fixed point can drive the late-time accelerated expansion. 

For $c=3$, the parameter space is a bit more restricted but still large enough to allows us recasting the eigenvalues in a similar fashion as before $(e_{1}, e_{2}, e_{3}, e_{4})$ with $e_{1}=0$. By examining figure (\ref{fig:2}) we infer the value $e_{2}=-3$ for $\lambda>0$ and $B_{0}>0$, or $e_{2}=-4$ for any value of $\lambda$ and $B_{0}<0$. On the other hand, $e_{3}=-3$ for $B_{0}>0$ and $\lambda<0$. Note that these two eigenvalues can be positive out of the inferred regions. Finally, $e_{4}=-3,-4$ for any value of $\lambda$. The former case is given by $B_{0}>0$ while the latter one by $B_{0}<0$. After putting together all the constraints, we see that there is no allowed parameter space that leads to a stable solution, that is to say, all eigenvalues having negative real parts. Moreover, their real parts are not all simultaneously positive (repeller) either. As a consequence, this fixed point has at least one eigenvalue with distinct sign, and can be classified as a \textit{saddle point}. It means that some trajectories in phase space pass close to this point but never end up here as required for a stable point. This fixed point hence corresponds to a \textit{vector-dark matter scaling solution}.

On the other hand, it is interesting to see that physical parameters depend on both the disformal coupling constant $B_{0}$ and the parameter $\lambda$. Even though they do not explicitly enter into the equation of $w_{\rm eff}$, they are decisive in setting the stability conditions as previously discussed. As a final remark $\Omega_{\rm DM}=2cB_{0}|\lambda|$ is a positive definite quantity consistent with the condition of existence such that $\Omega_{\rm DE}=1-2cB_{0}|\lambda|$ is always less than one, feature that is similar to the one found for the fixed point ($\rm H$). 

\item \textbf{Point} ($\rm G$): this solution corresponds to the case $\beta=-1/2$. This fixed point is a \textit{saddle point} with eigenvalues 
$(3/2, 3/2, -1, 0)$. It is characterized by matter domination $\Omega_{\rm DM}=1$ with presence of the vector field $v_{c}=\pm\frac{2}{B_{0}}$ but with no contribution to the content energy of the universe in the form of dark energy $\Omega_{\rm DE}=0$. Despite the presence of the vector field this solution does not represent properly a scaling solution. We shall see however that the presence of the vector field can make things a bit different in comparison to the standard matter domination epoch due to the disformal coupling.

\item \textbf{Point} ($\tilde{\rm G}$): this point is a \textit{saddle point} with $\beta=-2$ and characterized also by matter domination $\Omega_{\rm DM}=1$ with a presence of the vector field in a renewed form $v_{c}=\pm\sqrt{cB_{0}}$.  We parameterize the two different emerging solutions in terms of the same constant $c$, as was done above for the point ($\tilde{\rm H}$), but they have indistinguishable eigenvalues $(3/2, 3/2, -1, 0)$. This point, like ($\rm G$), is analogous to the ($\rm B$) point of the conformal case, sharing thus the same physical meaning.

\end{itemize}
In short, we have investigated the effects of both conformal and disformal couplings on the physical phase space by means of dynamical system analysis. Several critical points exist depending on the coupling-type, enriching thus the suitable phase space compared to the uncoupled case. In particular we have found scaling solutions where the vector field does not vanish during matter domination, and attractor points driving the late-time accelerated expansion different (or equal) to de Sitter solution. It remains to investigate, however, how they impact quantitatively the overall dynamics of the universe in comparison to the uncoupled case. This issue is treated by numerical methods in the following.

\section{Numerical results: Cosmological background evolution}\label{sec:5}
We have gained so far valuable information about the suitable parameter space from the dynamical system perspective that makes the present model cosmologically appealing to the light of current observations. In this regard, this analysis has served to examine the conditions under which the conformal and disformal couplings can provide stable cosmological solutions, such as scaling  attractor solutions that account for the current accelerated period. It is necessary for the purpose of better comprehension of how this can be visualized in a more realistic way to solve numerically the coupled system of equations and, thus, to verify all the qualitative features found. This aspect is explored separately in the next part for each coupling type.  

Overall, to study the background cosmological dynamics in these models, we integrate numerically the coupled system eqn.~(\ref{sec4:eqn3}), excluding the associated differential equations for the dynamical variables $x$ and $r$ due to the constraints $x^2=2\lambda v^2 y^2$ and the one given by eqn.~(\ref{sec4:eqn2}) that help us to reduce the number of differential equations. So, we are left with 4 differential equations that govern the evolution of the variables $u$, $v$, $y$ and $z$. In all the numerical computations, we set different sets of initial conditions at $N=-12$, well within the deep radiation-dominated era, such that these values lead to a consistent cosmological evolution\footnote{It worthwhile emphasizing that all initial conditions satisfy the constraint equation (\ref{sec3:eqn6}), which can be written in terms of the dynamical variables as $\frac{2 \lambda v^{2}y^{2}+2 u^{2}}{z^{2}}=g(C,C_{X},B,B_{X})$, with $g$ being a general function of the metric functions. This helps us, therefore, to control the  initial conditions taken in the numerical solutions. For instance, once the function $g$ is specified and for certain initial conditions of $v$, $y$ and $z$, the initial value of $u$ is completely determined.}. These are labeled with the superscript (i). In addition, we take initial conditions such that the energy density parameters match the present values ($N=0$) \cite{Planck:2018vyg}: $\Omega_{\rm DE}^{(0)}=0.68$ and $\Omega_{\rm r}^{(0)}\approx 1\times10^{-4}$. To do so, we implement the trial and error method as a recursive procedure to match the present values by adapting carefully the initial conditions. In particular, changes in the initial conditions of the variables $y$ and $u$ impact more notoriously the background dynamics, so these are allowed to vary mainly until success. Likewise,  we consider respectively the fiducial values $\lambda=-0.4$ and $v^{(i)}=0.11$, for the potential parameter and the normalized vector field, unless otherwise stated.  The numerical solution for the $\Lambda$CDM cosmological model is also shown in the plots for comparison. It helps us to understand better how these coupled models work at the background level.

Let us do a final remark on the initial conditions. We remind that $\Omega_{\rm DE}=x^2+y^2+u^2$ and $x^2=2\lambda v^2 y^2$, so if $v$ is taken to be large enough such that it compensates the small value chosen of $y$ as demanded by consistent cosmological solutions, $x$ can then contribute significantly to the energy density parameter of the vector field. This is not the case however because once $\lambda$ is fixed, $v$ is completely constrained according to eqn.~(\ref{sec4:eqn8}) to allow the existence of the attractor solution ($\rm D$). This is the reason why, even though $v$ does not vanish, its effect is almost negligible on the initial conditions. A very different situation is presented for the conformally and disformally attractor solutions where $v$ is quite less constrained (see tables (\ref{table1}) and (\ref{table2})) and can impact more visibly the cosmological solutions from the initial conditions as has been checked. This is not, however, for the purpose of this paper to explore the entire window of initial conditions. Our main concern is to investigate the effect of changing the values of the coupling parameters over a suitable cosmological evolution.

\begin{figure*}
\centering
\includegraphics[width=0.45\hsize,clip]{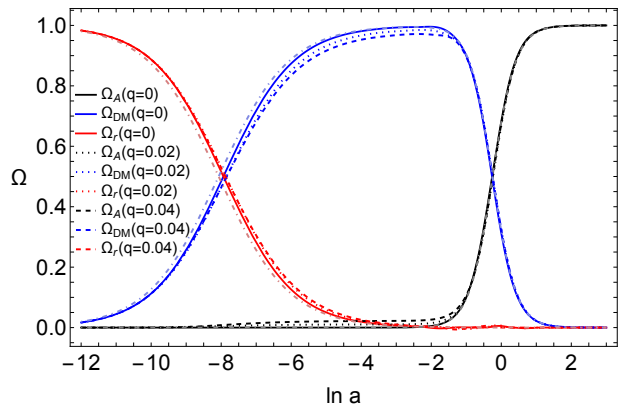}
\caption{Evolution of the density parameters 
 versus the number of e-folds $N=\ln{a}$ for different values of the conformal parameter $q$, describing the strength of the \textit{power law coupling} as denoted in the legend. Here $q=0$ (solid curves) represents the uncoupled case which is still different, at the background level, to the $\Lambda$CDM cosmological model (light dot-dashed curves) before fully matter domination. For each numerical computation we have taken the following initial conditions: for $q=0$, $u^{(i)}= 2\times10^{-10}, y^{(i)}= 2.95\times10^{-9}$; for $q=0.02$, $u^{(i)}= 1.3\times10^{-2}, y^{(i)}= 2.42\times10^{-9}$; for $q=0.04$, $u^{(i)}= 1.9\times10^{-2}, y^{(i)}= 2.1\times10^{-9}$. For all cases, we have chosen $v^{(i)}=0.11$ and $z^{(i)}= 1.3\times10^{-1}$ as fidutial values. All initial conditions have been chosen to match approximately the present values $\Omega_{\rm DE}^{(0)}=0.68$ and $\Omega_{\rm r}^{(0)}\approx1\times10^{-4}$.}\label{fig:3}
\end{figure*}
\begin{figure*}
\centering
\includegraphics[width=0.45\hsize,clip]{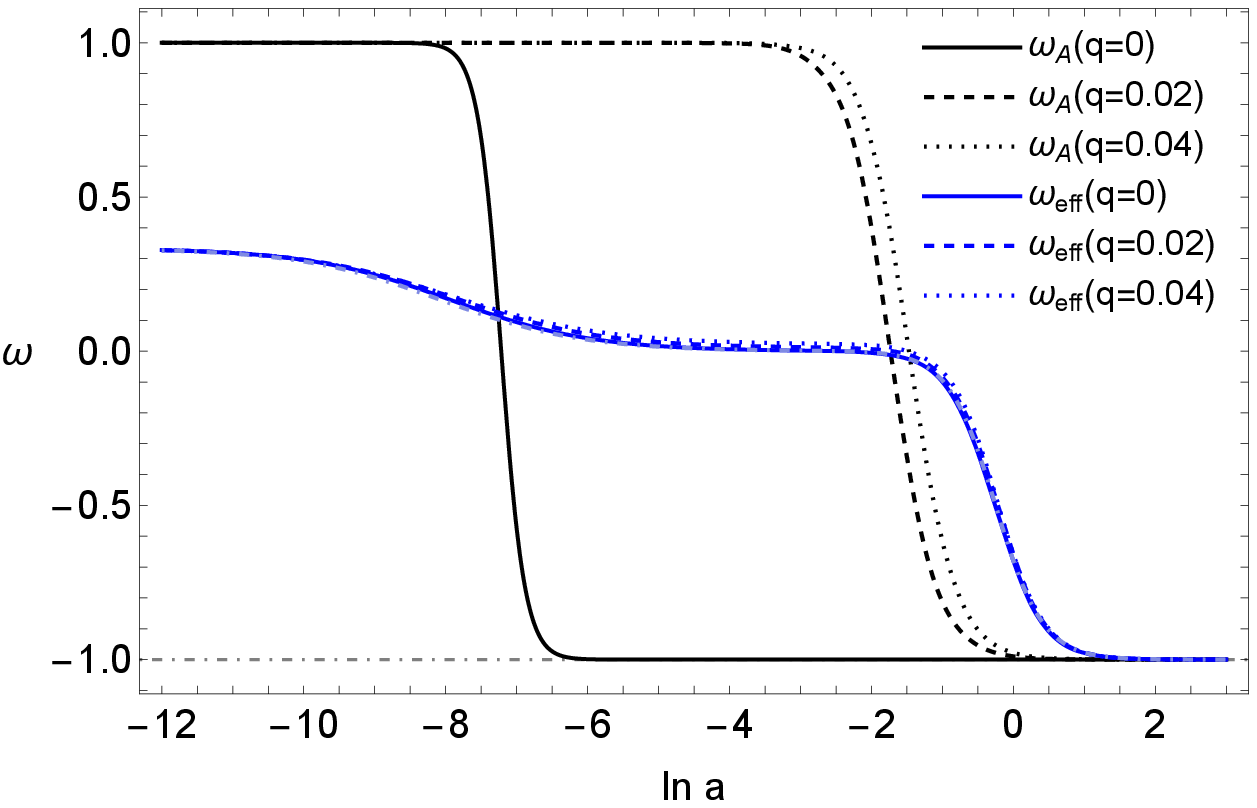}
\includegraphics[width=0.45\hsize,clip]{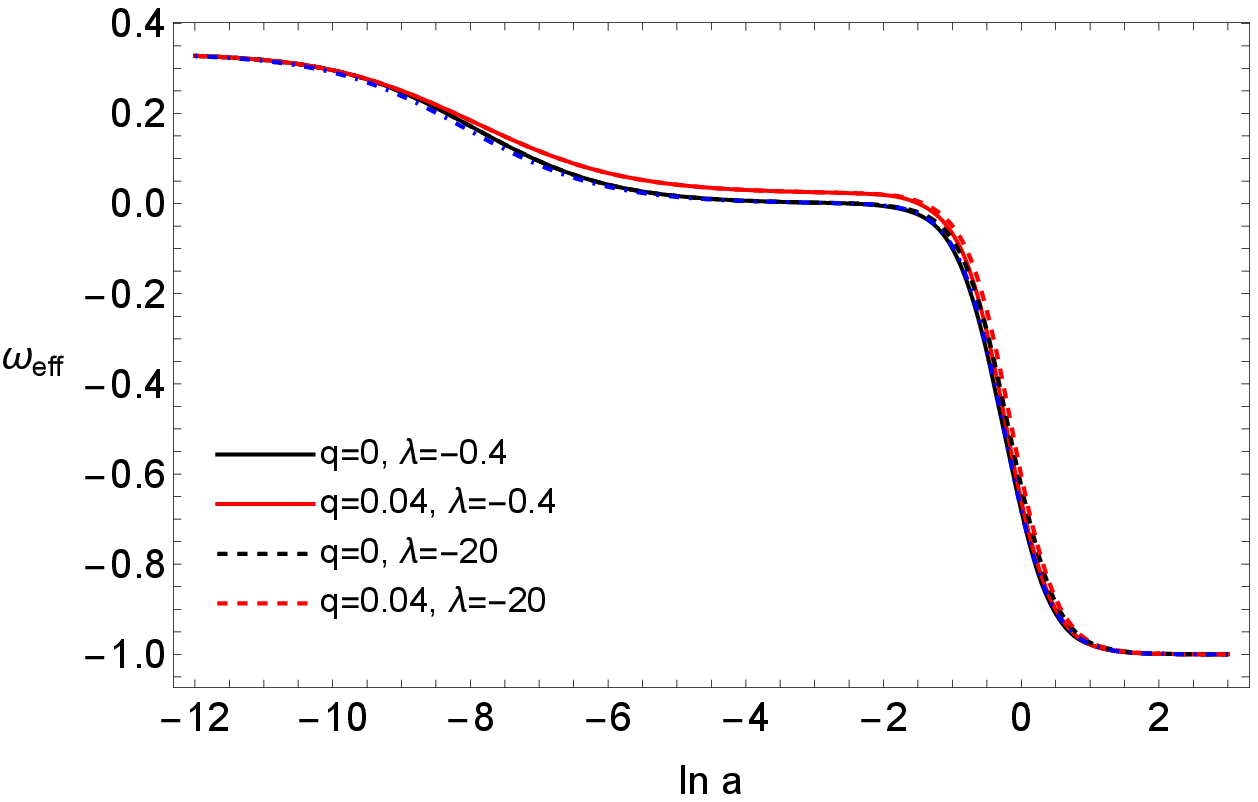}
\caption{Effective equation of state $w_{\rm eff}$ and equation of state for the vector field $w_{A}$ for different values of the model parameter, describing the uncoupled ($q=0$) and the conformally power law coupled case ($q\neq0$). Left panel depicts numerical solutions with $\lambda=-0.4$ and different values of $q$, as shown in the legend, for the same initial conditions as Figure (\ref{fig:3}). The cosmological model $\Lambda$CDM has been also included for comparison purposes (light dot-dashed curves). Right panel, instead, shows numerical solutions of the effective equation of sate only for two different values of $\lambda$ with associate $q$ values, the latter describing the uncoupled and coupled cases as indicated in the legend. As to the initial conditions, we have taken for $q=0$ and different $\lambda$, $u^{(i)}= 2\times10^{-10}, y^{(i)}= 2.95\times10^{-9}$ and $q\neq0$ and same $\lambda$,  $u^{(i)}= 2\times10^{-4}, y^{(i)}= 2.11\times10^{-9}$. Here the cosmological model $\Lambda$CDM is described by the blue dot-dashed line. As before, we have chosen for all cases, $v^{(i)}=0.11$ and $z^{(i)}= 1.3\times10^{-1}$ as fidutial values.}\label{fig:3a}
\end{figure*}

\subsection{Conformal case}
For the conformal case two specific coupling functions have been studied to figure out their effect in the cosmological dynamics, both impacting notoriously the evolution of the universe at different stages due to the appearance of novel critical points. The power law coupling, in particular,  supports the emergence of a vector-dark matter scaling solution ($\rm C$) that may affect the evolution of structures in a different way when comparing to the standard $\Lambda$CDM scenario and also to the uncoupled case. The first distinction  one notices at once is that dark matter does not dominate fully the content energy of the universe but, instead, there is a novel contribution of the vector field in the form of dark energy due to the coupling $q$. Specifically,  $\Omega_{\rm DM}=\frac{-2+6q}{-2+5q}$ and $\Omega_{\rm DE}=\frac{q}{2-5q}$. It implies that as long as $q$ increases, $\Omega_{\rm DM}$ comes down while $\Omega_{\rm DE}$ grows. To check this feature we perform some numerical computations for different values of the parameter $q$ within the allowed range ($0<q<1/3$) and plot the evolution of the energy density parameters in figure (\ref{fig:3}). This feature is clearly evidenced as $q$ increases. As a merely qualitative aspect, the radiation-matter equality is slightly shifted as $q$ increases as well, it happening later compared to the uncoupled case $(q=0)$ and the $\Lambda$CDM cosmological model (light dot-dashed curves). For the coupled cases, it is also observed an early onset of the growth of dark energy around radiation-matter equality  (see dotted and dashed curves).

It is also instructive to see the evolution of the equation of state of the vector field and how it tracks the effective equation of state parameter at late times. This is depicted in the left panel of figure (\ref{fig:3a}). The vector field behaves as a stiff fluid at early times ($w_{A}=1$) and as dark energy ($w_{A}=-1$) either shortly after radiation-matter equality or just at the present epoch, depending on the coupling parameter. In the right panel of the same figure the effect of changing both $q$ and $\lambda$ is shown for the sake of completeness. Varying $\lambda$, for instance, may affect the late time evolution of the universe as expected since the potential energy plays the role of dark energy. This can be appreciated in the plot because there are no differences between the solid black and dashed black curves, which have same $q$ and different $\lambda$ values, during radiation and matter dominations. Here, different colors stand for different values of $q$ and same $\lambda$. Conversely, the effect of changing $\lambda$ is visible at late times. It is interesting to see, on the other hand, that changes in the parameter $q$ are distinguishable in most of the evolution of the universe, leading to an interaction term $\gamma_{C}\sim\mathcal{O}(10^{-2})$ during all the cosmic evolution. 

\begin{figure*}
\centering
\includegraphics[width=0.45\hsize,clip]{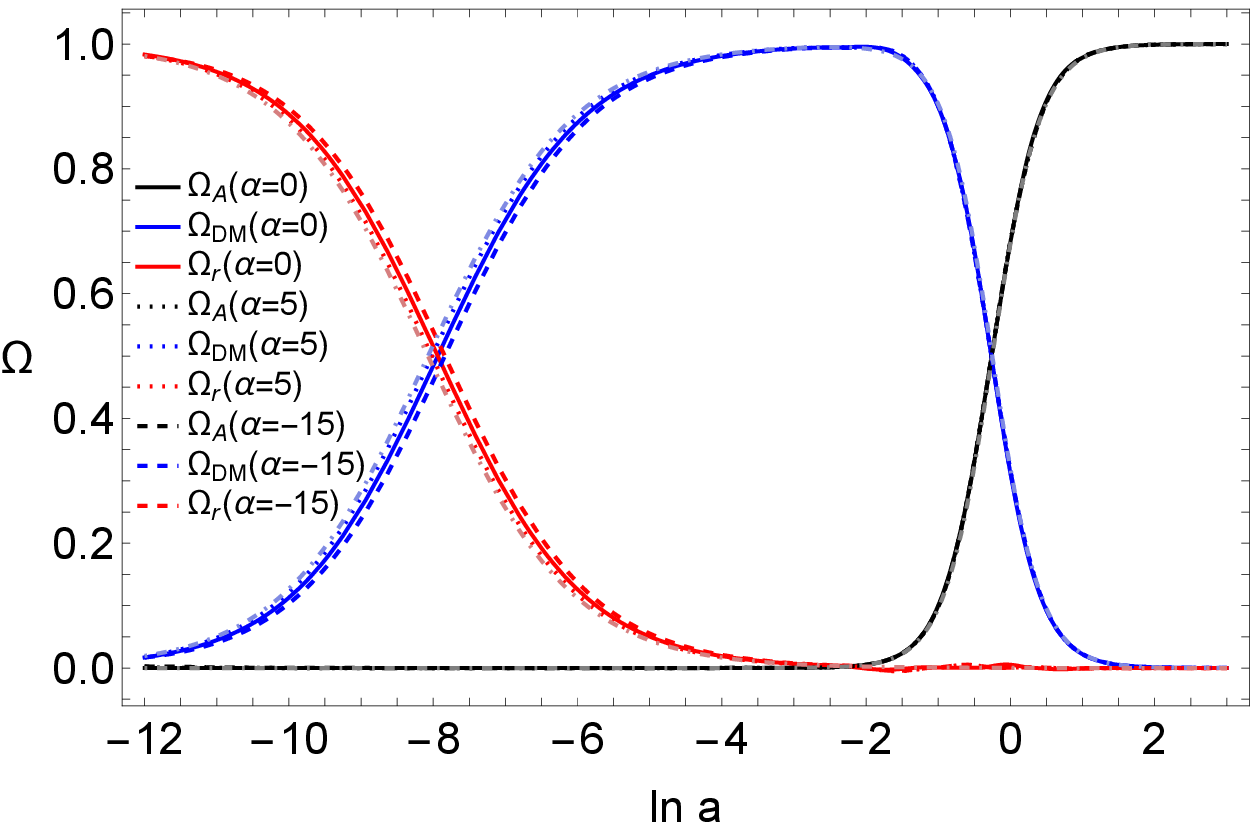}
\includegraphics[width=0.45\hsize,clip]{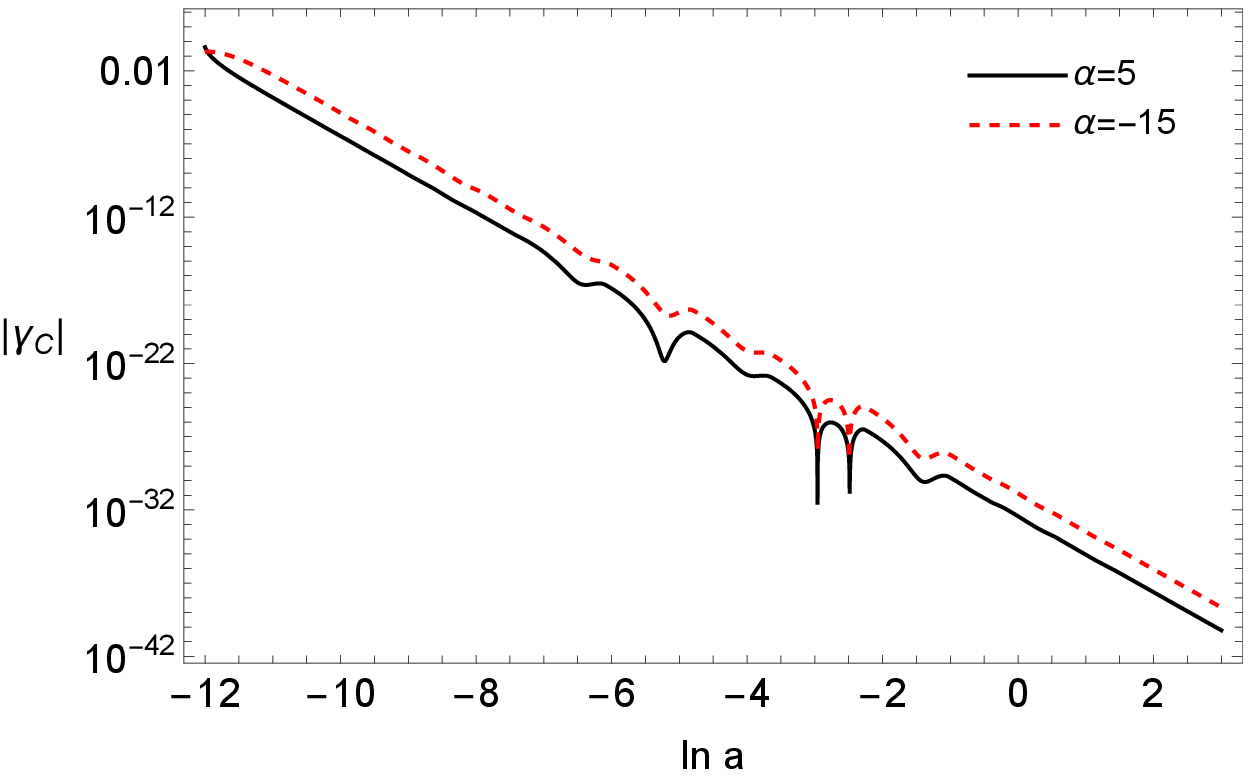}
\caption{Left panel shows the evolution of the density parameters  versus the number of e-folds $N=\ln{a}$ for different values of the conformal parameter $\alpha$, describing the strength of the \textit{exponential coupling} as denoted in the legend. Here $\alpha=0$ represents the uncoupled case which is appreciably different to the $\Lambda$CDM cosmological model (light dot-dashed curves) before fully matter domination. For each numerical computation we have taken the following initial conditions: for $\alpha=0$, $u^{(i)}= 2\times10^{-10}, y^{(i)}= 2.95\times10^{-9}$; for $\alpha=5$, $u^{(i)}= 5\times10^{-2}, y^{(i)}= 3.05\times10^{-9}$; for $\alpha=-15$, $u^{(i)}= 5\times10^{-2}, y^{(i)}= 2.79\times10^{-9}$. For all cases, we have chosen $v^{(i)}=0.11$ and $z^{(i)}= 1.3\times10^{-1}$ as fidutial values. All initial conditions have been chosen to match approximately the present values $\Omega_{\rm DE}^{(0)}=0.68$ and $\Omega_{\rm r}^{(0)}\approx1\times10^{-4}$. Right panel shows the evolution of the absolute value of the interaction term for the exponential coupling for the same initial conditions as left panel. Here the effect of changing the sign of $\alpha$ over the strength of $|\gamma_{c}|$ is assessed as well.}\label{fig:3b}
\end{figure*}
\begin{figure*}
\centering
\includegraphics[width=0.45\hsize,clip]{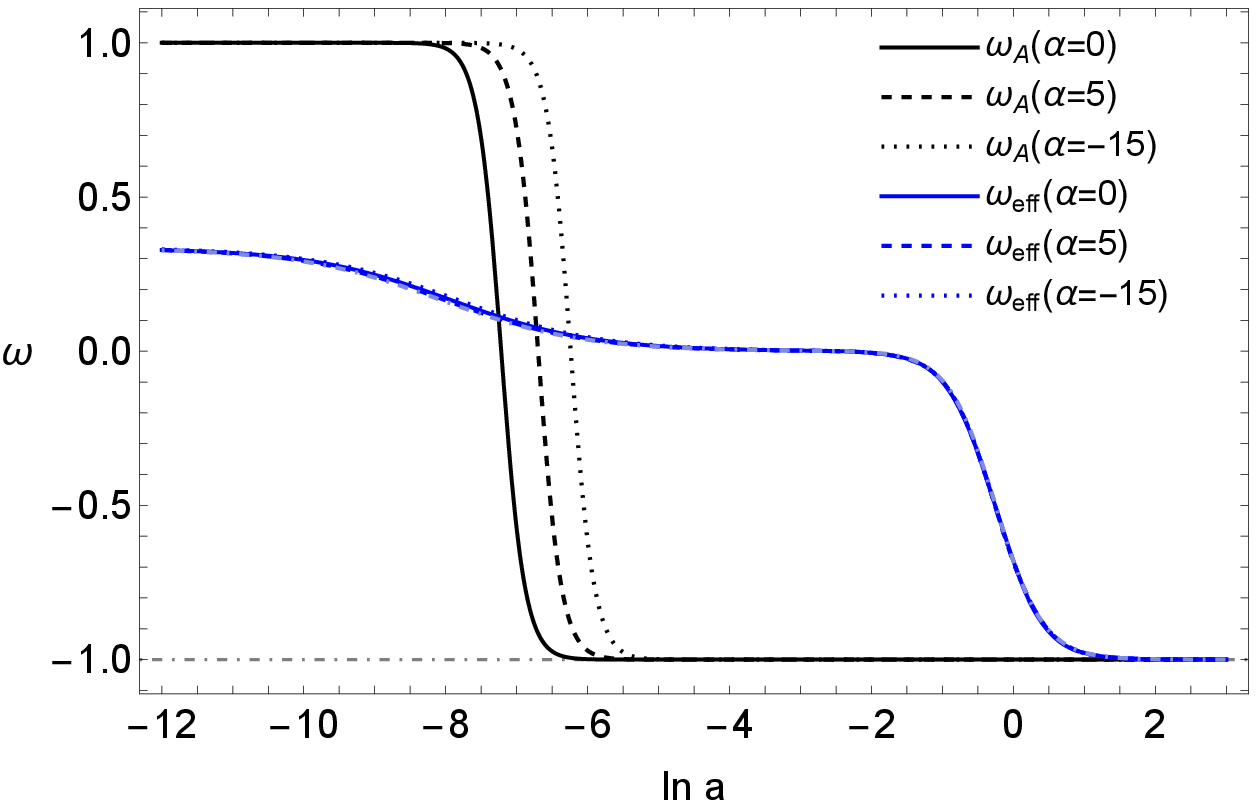}
\includegraphics[width=0.45\hsize,clip]{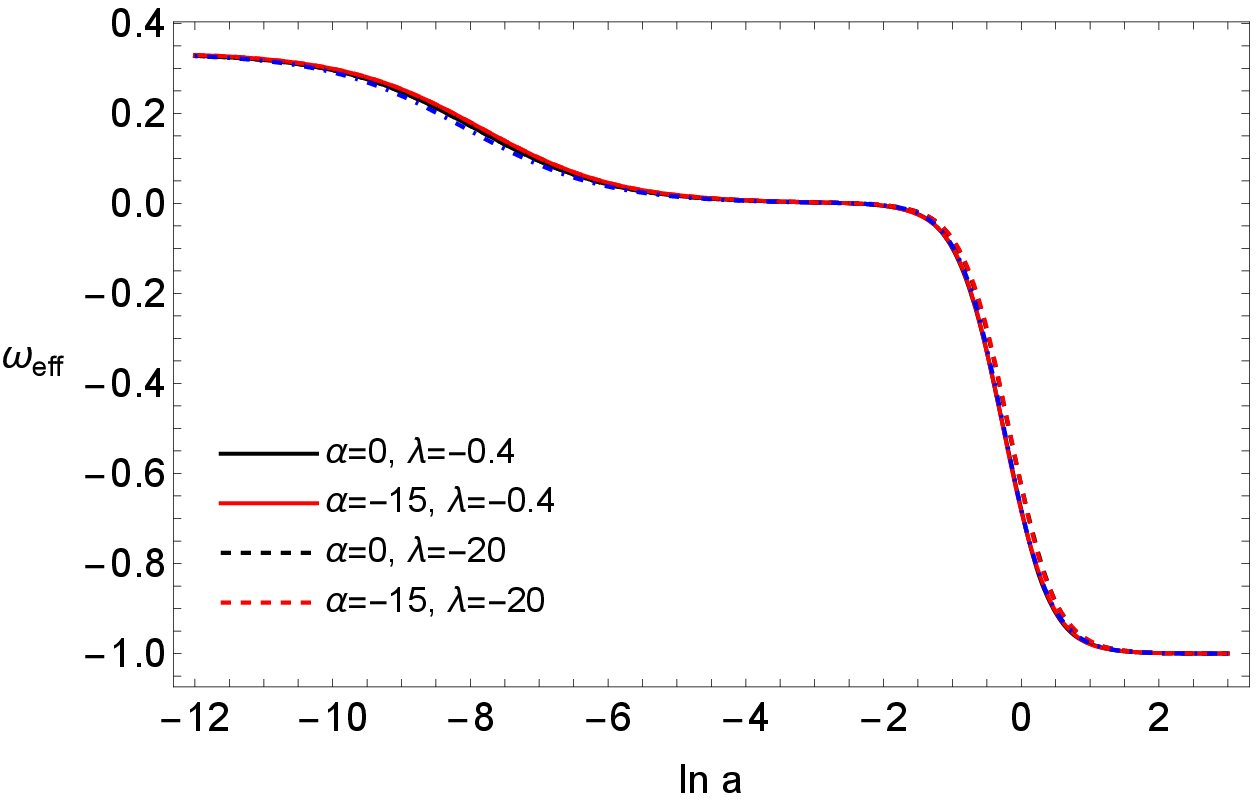}
\caption{Effective equation of state $w_{\rm eff}$ and equation of state for the vector field $w_{A}$ for different values of the model parameter, describing the uncoupled ($\alpha=0$) and the conformally exponential coupled case ($\alpha\neq0$). Left panel depicts numerical solutions with $\lambda=-0.4$ and different values of $\alpha$, as shown in the legend, for the same initial conditions as figure (\ref{fig:3b}). The cosmological model $\Lambda$CDM has been also included for comparison purposes (light dot-dashed curves). Right panel, instead, shows numerical solutions of the effective equation of sate only for two different values of $\lambda$ with associate $\alpha$ values, the latter describing the uncoupled and coupled cases as indicated in the legend. As to the initial conditions, we have taken for $\alpha=0$ and different $\lambda$, $u^{(i)}= 2\times10^{-10}, y^{(i)}= 2.95\times10^{-9}$ and $\alpha\neq0$ and  same $\lambda$,  $u^{(i)}= 5\times10^{-2}, y^{(i)}= 2.79\times10^{-9}$. Here the cosmological model $\Lambda$CDM is described by the blue dot-dashed line. As before, we have chosen for all cases, $v^{(i)}=0.11$ and $z^{(i)}= 1.3\times10^{-1}$ as fidutial values.}\label{fig:3c}
\end{figure*}

As to the exponential coupling case, its dynamical character is mostly encoded in the critical points ($\rm \tilde{D}$) and ($\rm \tilde{F}$) with the former describing the current accelerated expansion. 
One interesting fact is the appearance of both $\alpha$ and $\lambda$ parameters in the physical quantities, in contrast to the power law coupling, as found from the dynamical system analysis; $\alpha$ being the parameter that accounts for the strength of the coupling to dark matter and is the one we must pay much of our attention. Before exploring the cosmological dynamics, note that $\alpha=0$ is excluded by demanding the existence of the critical points ($\rm \tilde{D}$) and ($\rm \tilde{F}$). It means that taking $\alpha=0$ leads to another trajectory in phase space given by the uncoupled solution $(\rm D)$. Therefore, two different trajectories can exist depending on $\alpha$. One can immediately ask how different they are from each other and also from the power law coupling. This is investigated by computing the evolution of the energy density parameters of all components considered and showed in the left panel of figure (\ref{fig:3b}). The uncoupled case is identified as solid curves and the coupled ones as all other types of curves as can be read from the legend. The $\Lambda$CDM cosmological model has been included for comparison and described by light dot-dashed curves. Likewise positive and negatives values of $\alpha$ are allowed and then explored here for the available parameter space. Note that the sign of $\alpha$ has the effect of increasing ($\alpha>0$) or reducing ($\alpha<0$) the energy density parameter of dark matter ($\Omega_{\rm DM}= \frac{|\lambda|}{\alpha}$)
at the expense of dark energy $\Omega_{A}=1-\frac{|\lambda|}{\alpha}$, this latter being 
practically unaffected before matter domination. They are seen above or below the solid curves as appropriated. Nevertheless, the most visible effect of changing $\alpha$ is around radiation-matter equality, with no distinguishable features at late times even when compared to the uncoupled case. Hence the coupling effect is important only before fully dark matter domination. In the deep radiation dominated era the coupling naturally turns off.

In the right panel of the same figure the (absolute value of) interacting term is plotted, which accounts for the strength of the vector coupling to dark matter. Here the cusps represent the change of sign of the vector field $v$ (and not of $\gamma_{c}$). Notice also that $\alpha>0$ provides $\gamma_{c}<0$ and vice versa. At very early times $|\gamma_{c}|\sim\mathcal{O}(0.1)$, but it decays fastly to very small values today, leaving a very narrow room to look for differences between the uncoupled case. This is, indeed, the reason why all the numerical solutions for the energy density parameters exhibit small differences with respect to the uncoupled case as strongly suspected, fact that becomes now more transparent. It indicates also, as a direct consequence, that fits high-redshift data (like BBN and CMB temperature anisotropies) may be more sensible to the coupling effects than low-redshift data whereby implementation of cosmological data at high redshift is a promising way to proceed in order to constrain the conformally coupled exponential model.

Also, the effective equation of state parameter and the vector field equation of state are shown in the left panel of figure (\ref{fig:3c}) for different values of $\alpha$ as before. Similar to the power law coupling case, the vector field equation of state is more noticeably affected as the coupling parameter $\alpha$ changes (see left panel) according to $w_{A}=-\frac{1}{1+\frac{\lambda}{\alpha}}$. An interesting difference between the power law coupling and the present case is the fact that taking $\alpha>0$, the vector field equation of state changes sooner from stiff fluid ($w_{A}=1$) to dark energy ($w_{A}=-1$). Another feature seen in the right panel of figure (\ref{fig:3c}) is that $\alpha$ has almost negligible impact on the effective equation of state compared to the uncoupled case: different colors which stand for different $\alpha$ are indistinguishable. There is however visible differences with respect to the $\Lambda$CDM cosmological model in the early universe that would be worth quantifying with the help of observational data in order to assess the cosmological viability of the model. Thus, we have investigated so far some cosmological consequences of the conformally coupled models and their qualitative differences at the background level for the available parameter space based on numerical analysis. It remains to compute numerically the cosmological evolution of the disformally coupled model in order to investigate the effects of the coupling parameters on the background dynamics.

\begin{figure*}
\centering
\includegraphics[width=0.45\hsize,clip]{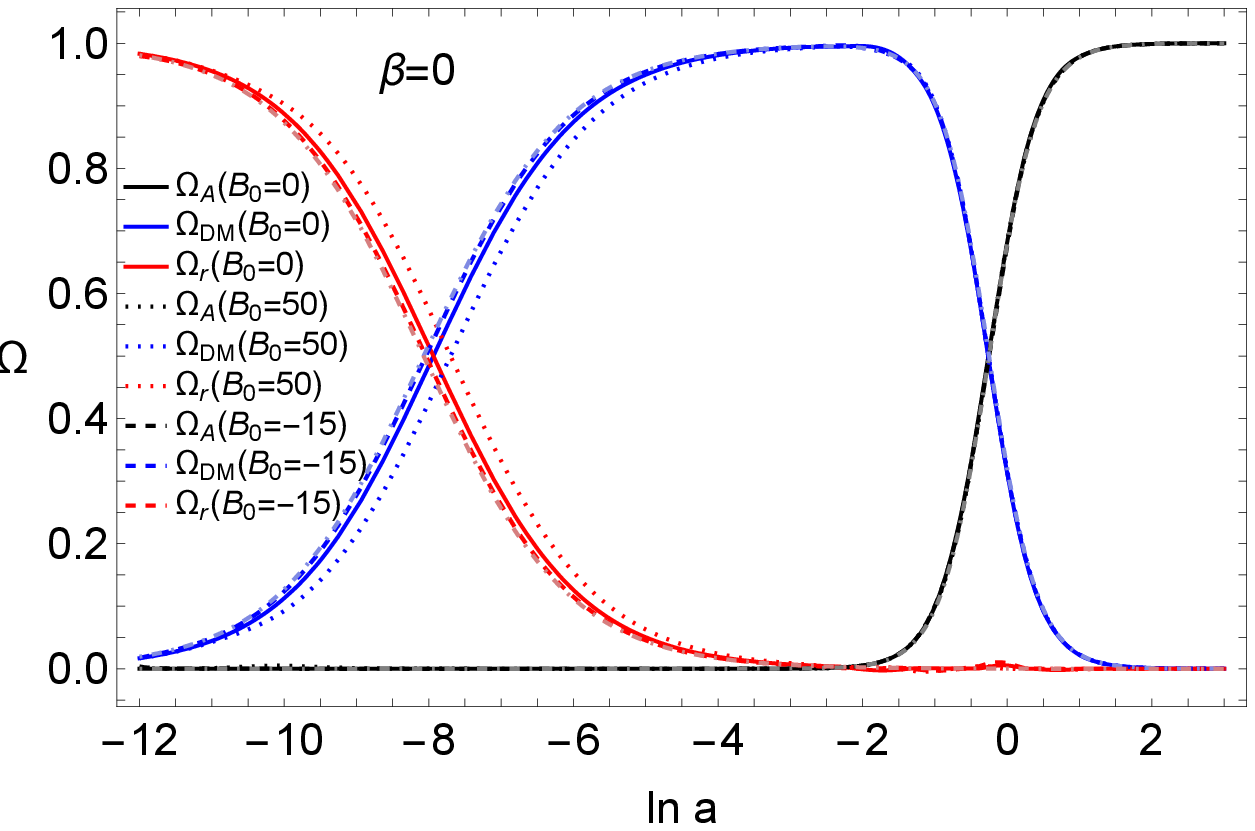}
\includegraphics[width=0.45\hsize,clip]{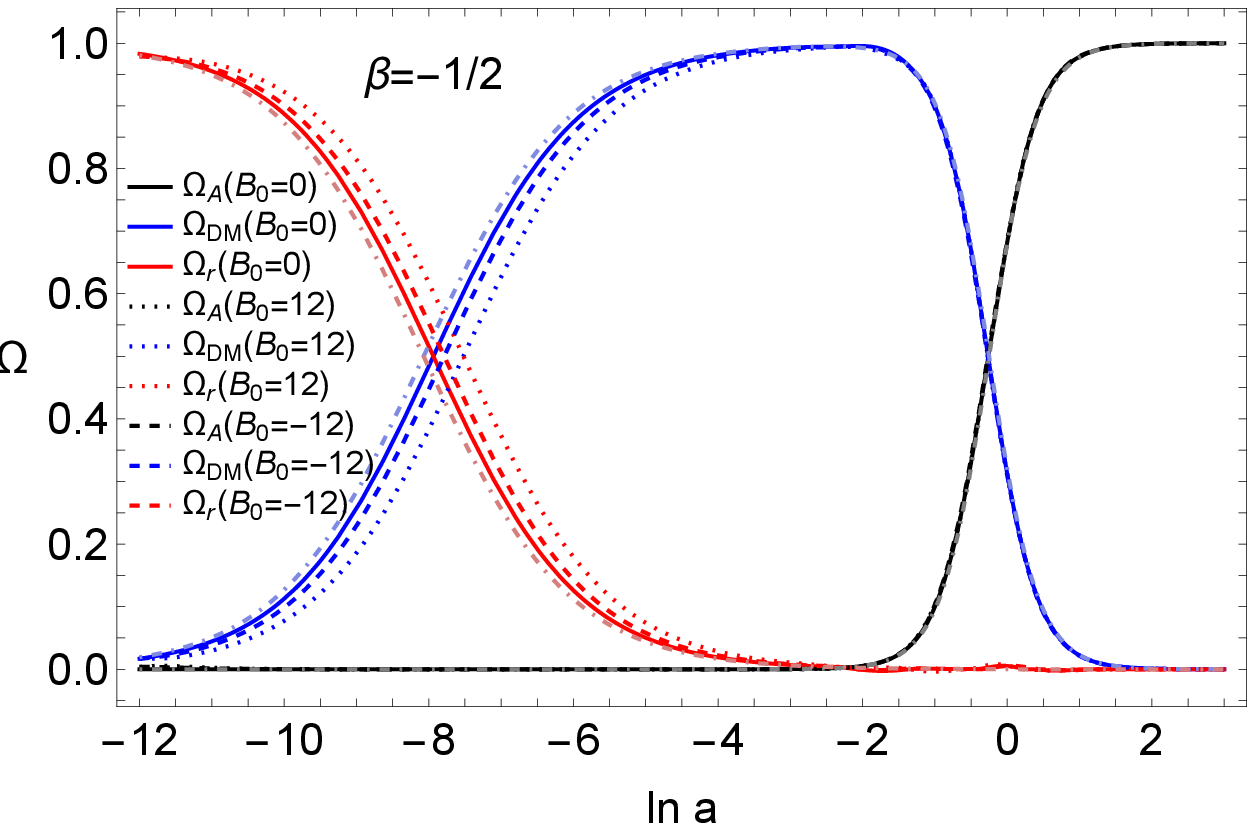}
\caption{Evolution of the density parameters 
 versus the number of e-folds $N=\ln{a}$ for different values of the disformal parameter $B_{0}$ as described in the legend. Left panel corresponds to the case $\beta=0$ with the following initial conditions associated to each numerical computation: for $B_{0}=0$, $u^{(i)}= 2\times10^{-10}, y^{(i)}= 2.95\times10^{-9}$; for $B_{0}=50$, $u^{(i)}= 5\times10^{-3}, y^{(i)}= 2.59\times10^{-9}$; for $B_{0}=-15$, $u^{(i)}= 6.1\times10^{-2}, y^{(i)}= 3.08\times10^{-9}$. Right panel shows numerical solutions for the power law $\beta=-1/2$ where the attractor-like solution was found. The following initial conditions have been taken: for $B_{0}=12$, $u^{(i)}= 6.1\times10^{-2}, y^{(i)}= 2.4\times10^{-9}$; for $B_{0}=-12$, $u^{(i)}= 6.1\times10^{-2}, y^{(i)}= 2.76\times10^{-9}$. In both cases the solution $B_{0}=0$ represents the uncoupled case which is analogous to conventional quintessence models and appreciably different to the $\Lambda$CDM cosmological model (see light dot-dashed curves) before fully matter domination. For all cases, we have chosen $v^{(i)}=0.11$ and $z^{(i)}= 1.3\times10^{-1}$ as fidutial values. All initial conditions have been chosen to match approximately the present values $\Omega_{\rm DE}^{(0)}=0.68$ and $\Omega_{\rm r}^{(0)}\approx1\times10^{-4}$.}\label{fig:5}
\end{figure*}
\begin{figure*}
\centering
\includegraphics[width=0.45\hsize,clip]{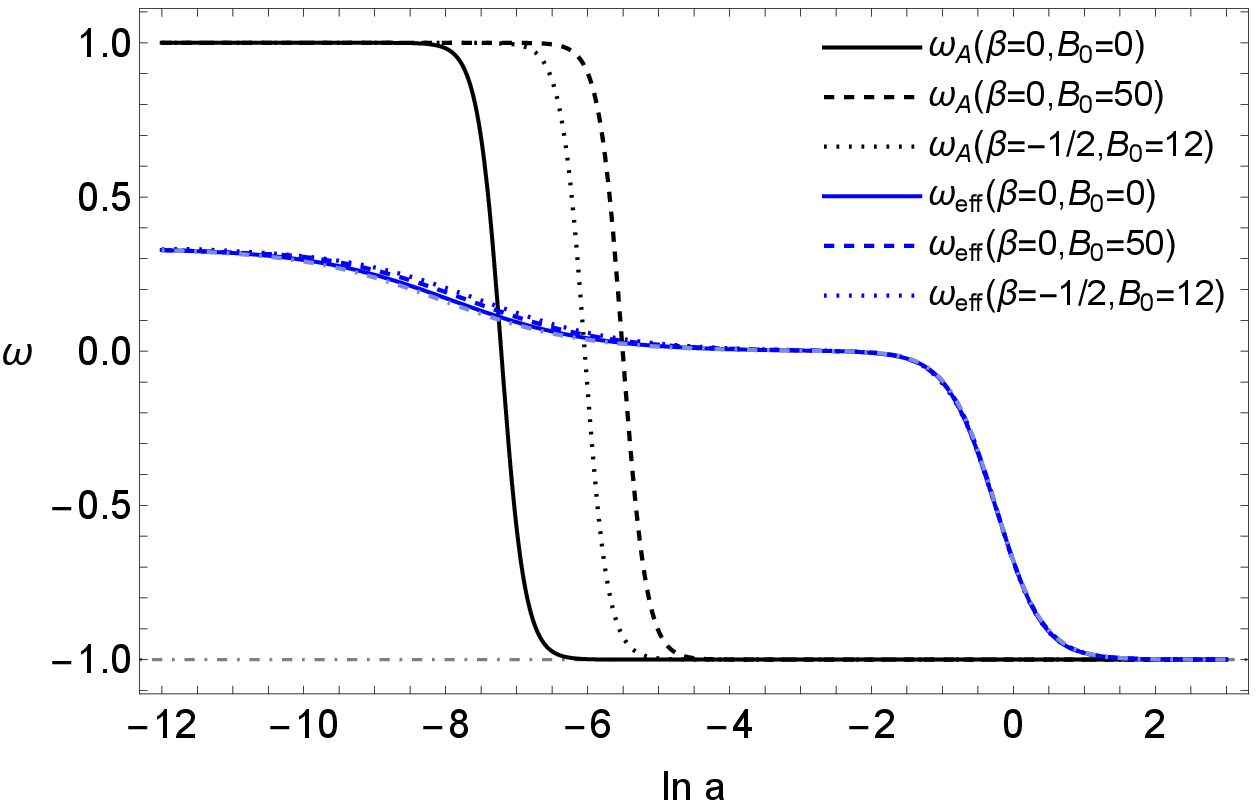}
\includegraphics[width=0.45\hsize,clip]{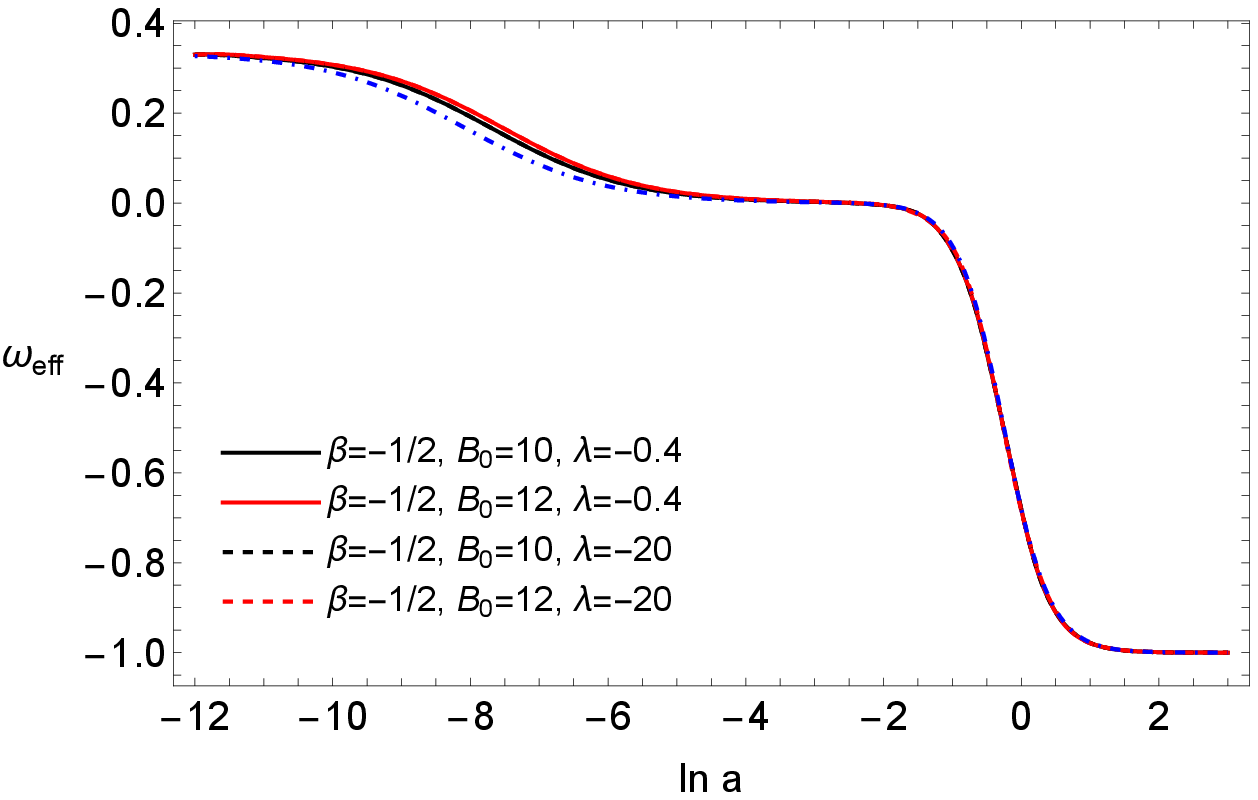}
\caption{Effective equation of state $w_{\rm eff}$ and equation of state for the vector field $w_{A}$ for different values of the model parameter, describing the uncoupled ($B_{0}=0$) and disformally coupled cases ($B_{0}\neq0$). Left panel depicts numerical solutions with $\lambda=-0.4$ and different values of the coupling parameters $\beta_{0}$ and $B_{0}$, as shown in the legend, for the same initial conditions as figure (\ref{fig:5}). The cosmological model $\Lambda$CDM has been also included for comparison purposes (see light dot-dashed curves). Right panel, instead, shows numerical solutions of the effective equation of state only for two different values of $\lambda$ and $\beta=-1/2$. For the case $\lambda=-0.4$ and $B_{0}=12$ the same initial conditions as the right panel of figure (\ref{fig:5}) have been taken. For the same $\lambda$ and $B_{0}=10$, the selected initial conditions are $u^{(i)}= 6.1\times10^{-2}, y^{(i)}= 2.63\times10^{-9}$. For $\lambda=-20$, the following initial conditions have been chosen: for $B_{0}=10$, $u^{(i)}= 5.9\times10^{-2}, y^{(i)}= 2.96\times10^{-9}$; for $B_{0}=12$, $u^{(i)}= 6\times10^{-2}, y^{(i)}= 2.68\times10^{-9}$. Here the cosmological model $\Lambda$CDM is described by the blue dot-dashed line. As before, we have chosen for all cases, $v^{(i)}=0.11$ and $z^{(i)}= 1.3\times10^{-1}$ as fidutial values.}\label{fig:6}
\end{figure*}
\begin{figure*}
\centering
\includegraphics[width=0.45\hsize,clip]{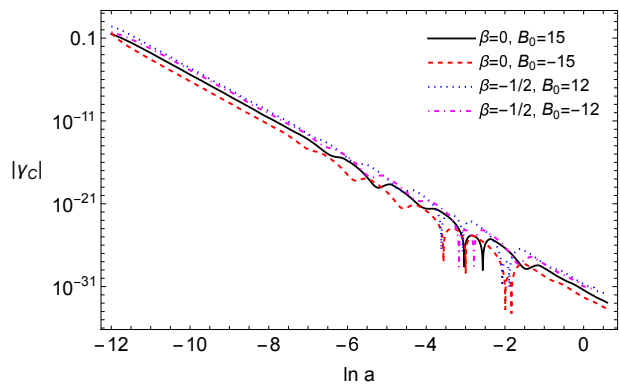}
\caption{Evolution of the absolute value of the interaction term for different values of the disformal model parameters as indicated in the legend. Here the effect of changing the sign of $B_{0}$ over the strength of $|\gamma_{c}|$ is assessed. All numerical computations correspond to $\lambda=-0.4$ with the same initial conditions as figure (\ref{fig:5}) and (\ref{fig:6}) as appropriate. For $B_{0}=15$, which has not been included before in our analysis, the same initial conditions as for $B_{0}=-15$ are chosen.}\label{fig:7}
\end{figure*}

\subsection{Disformal case}
We start by plotting the energy density parameters for the allowed region of the parameter space in accordance with the dynamical system analysis performed in section \ref{sec:4.1.2}. As anticipated, even thought the case with $\beta=0$ does not lead to new critical points in comparison with the uncoupled case, this constant coupling case can impact notoriously the cosmological background evolution in a non-trivial way because its associated differential equations have a more involved global structure.  Numerical solutions are depicted in the left panel of figure (\ref{fig:5}) for different values of the disformal parameter $B_{0}$. Solid curves represent the cosmological evolution for the uncoupled case $B_{0}=0$. Deviation of this solution are given by dotted and dashed curves due to the disformal coupling with $B_{0}\neq0$ as described in the legend. As in the conformally exponential case, the energy density parameters can be reduced or increased depending on the sign of $B_{0}$, with the difference that here negative coupling parameters increase instead the energy density of dark matter. 
We can see that numerical solutions can reproduce fairly well the entire cosmological dynamics with the expected transient periods: radiation, matter and dark energy dominations. Hence, whether $\beta=0$ is preferred over other possibilities, it is a subject that must be evaluated in the cosmological parameter estimation when calculating the best-fit parameters from observational data. This goes of course beyond the scope of the current study. So, this particular case can not be discarded \textit{a priori} at all from the analysis presented here.

In contrast, numerical solutions with $\beta=-1/2$ are plotted in the right panel of figure (\ref{fig:5}) which correspond to the stable attractor solution ($\rm H$), they being still dissimilar to the uncoupled case (solid curves with $B_{0}=0$). We perform numerical computations for different values of $B_{0}$ like the previous case ($\beta=0$). However, a key difference is that, no matter the sign of $B_{0}$, they always reduce the energy density of dark matter in comparison to the uncoupled case since $\Omega_{\rm DM}\propto 1/B_{0}^{2}$. These numerical solutions are clearly distinct from to the $\Lambda$CDM cosmological model (see light dot-dashed curves). We strongly suspect that differences between positive and negative choices of $B_{0}$ come from the interacting term (eqn.~(\ref{sec:4:eqn1})) where the sign really matters for the numerical solution. As $B_{0}>0$ leaves more visible imprints on the cosmological dynamics, we will focus mainly on this latter for the sake of illustration, without taking any prejudiced position over the negatives values. On the other hand, it is appreciable that the disformal coupling lowers the matter-radiation equality towards the present, as well as the energy density of dark matter, more appreciably in comparison to the case $\beta=0$. Furthermore, increasing $B_{0}$ and taking $\lambda$ fixed,
$\Omega_{\rm DE}$ gets bigger and $\Omega_{\rm DM}$ smaller since $\Omega_{\rm DE}=1-\frac{8|\lambda|}{B_{0}^{2}}$ and $\Omega_{\rm DM}=\frac{8|\lambda|}{B_{0}^{2}}$. Notice also that the saddle point $(\rm G)$, corresponding to matter domination period with $v_{c}=\pm\frac{2}{B_{0}}$, can be present here once the coupling is turned on. Changing for instance the initial condition for $v$, say, $10\%$ it strengths the interacting term (eqn.~(\ref{sec:4:eqn1})) almost a factor of two bigger for $B_{0}=15$ as we have checked. For smaller values of $B_{0}$ the effect of $v$ turns out to be less important. Interestingly, the presence of the vector field can affect the global cosmological evolution through the disformal coupling even when it does not contribute to the energy density in the form of dark energy as can be read from the numerical solutions.

Incidentally, we display in figure (\ref{fig:6}) the equation of state parameter for different values of the model parameters as indicated in the legend. From the left panel we can conclude that the vector field contributes to the energy density in the form of stiff fluid in the early universe (fixed point ($\rm E$)), while in the matter domination period, it begins to behave like dark energy, driving the accelerated expansion once it dominates the energy content of the universe. The precise time depends, in addition to $\lambda$, on the disformal coupling parameters. Also, we can see how $w_{A}$ tracks $w_{\rm eff}$ at very late times. Thus, it is very instructive to see how the effective equation of state follows the general and demanded trend with the transitions $w_{r}\to w_{m}\to w_{\rm DE}$, ensuring the radiation, matter, and dark energy periods as in the conformal cases. In the right panel of the same figure, the effect of varying $\lambda$ on the effective equation of state for $\beta=-1/2$ has been assessed. As expected, the effect of changing  $\lambda$, keeping the other model parameters fixed, may be relevant only once the energy density parameter of the vector field starts to evolve. This effect is not visible here because it is compensated with the fact of taking different initial conditions that must match the present energy density parameters as demanded. In contrast, the effect of changing slightly the value of the disformal coupling parameter $B_{0}$, keeping this time $\lambda$ fixed, is barely appreciable during radiation-matter equality: compare curves with different colors for either of the curve styles shown\footnote{Note however that  dashed curves are overlapping with their respective solid ones so that discrimination between $B_{0}=10$ (black dashed curve) and $B_{0}=12$ (red dashed curve) for $\lambda=-20$ is not appreciable.}.

Finally, we plot in figure (\ref{fig:7}) the evolution of the (absolute value of) interacting term for some cases studied previously. Here the interacting term can be positive or negative depending on $\beta$ and on the sign of $B_{0}$,\footnote{The red dashed curve represents indeed the only solution for which $\gamma_{c}<0$, corresponding to the choice $\beta=0$ and $B_{0}=-15$. According to the stability constraints found in section \ref{estability}, this solution must be however ruled out.} with $\gamma_{c}\sim\mathcal{O}(1)$ at very early times. However, they all also decay fastly to very small values today as the universe evolves similar to the exponential coupled case. Note that here numerical integration has been stopped around $N=0$ because some numerical solutions beyond this point are essentially zero for numerical precision purposes. 

As a main conclusion from the numerical analysis performed, the interacting term for the disformal coupling can be up to five order of magnitude larger than the one associated to the conformally exponential coupling during all the cosmological evolution. Nevertheless, the interacting term for the conformally power law coupling is larger today $\gamma_{C}\sim\mathcal{O}(10^{-2})=\rm const.$, and only a bit smaller at early times, though it quickly becomes larger as soon as the universe evolves.
This shows the rich possibility in exploring the effects of the conformal and disformal couplings at different stages of the evolution of the universe. Thus, we have assessed in a more quantitative manner the effects of the coupling parameters and their main differences by investigating their impact on the cosmological evolution.

\section{Discussion and conclusions}\label{sec:6}


Coupled dark energy models have brought the attention because of the rich phenomenology they can provide when contrasting with observational data. Thus, at the most phenomenological level these kind of scenarios can offer a promising alternative to solve some tensions revealed recently in the $\Lambda$CDM cosmological model. From the side of theoretical foundations one formal way to build interactions, at the level of the action, is by assuming that the dark matter sector is described by a metric that is related to the one of the gravitational part by a non-trivial disformal transformation that leaves the causal structure of spacetime unaltered. Following this conception, in most of the coupled dark energy models
the gravitational sector of the theory is of the scalar–tensor nature with the scalar field
playing the role of dark energy, and the coupling of the scalar field to dark matter is described
via either conformal or disformal transformations. A natural question that comes to our minds is, can vector fields identified as dark energy be coupled to dark matter and offer the same virtues as scalar fields do by conformal/disformal transformations? We have showed in this paper that this question can be answered favorably, putting thus, from a purely phenomenological perspective, vector fields in the same
privileged status that scalar fields occupy in coupled dark energy models. In the process of finding a convincing response some theoretical and numerical strategies have been used for the sake of completeness. We summarize here our main findings based on those approaches.

The resulting interaction term has been derived quite independent of the gravitational sector but demanding up to second-order derivative contributions sourcing the equations of motion eqn.~(\ref{sec2:eqn10}) through  eqn.~(\ref{sec2:eqn11}) to prevent the presence of Ostrogradski instabilities at this stage. This condition is however easily achieved because of the simple form of the vector disformal transformation eqn.~(\ref{sec2:eqn2}) that facilitates in turn all the analytical treatments.
Higher derivatives of the vector field can be also regarded in the conformal and disformal coupling functions but it requires integrating out the auxiliary degree of freedom to have second-order equations of motion. This alternative deserves to be explored as a theoretical possibility to generalize our results following the spirit of \cite{Gumrukcuoglu:2019ebp}. This result can also be applied to more general vector-tensor theories like the surviving part of the Generalized Proca theory ($\mathcal{L}_{3}$) and also to extended vector tensor theories. Interestingly, a more involved continuity equations have been obtained compared to the scalar counterpart with a novel coupling of the vector field to dark matter as evidenced in eqn.~(\ref{sec2:eqn14}). To put this in a concrete cosmological setup, the standard Proca theory minimally coupled to gravity with a vector (exponential) potential has been assumed to describe the gravitational sector. A direct consequence of this choice is that the non-propagating degree of freedom can address the accelerated expansion today in a FLRW universe, feature that is naturally reminiscent to the Generalized Proca theory and different from what is observed in the scalar field case. So, the interaction term provides new branches of solutions satisfying the equation of motion of the vector field in comparison to the uncoupled case where the trivial solution $A=0$ is allowed only (see eqn.~(\ref{sec3:eqn6})). This, of course, enriches the cosmological dynamics in an interesting way, depending, in turn, on the functional form assumed for the conformal and disformal couplings.

For the sake of concreteness, we have studied the cosmological dynamics of the coupled vector dark energy scenarios, assuming for the conformal coupling a power law and an exponential functions, and for the disformal case a general power law.  Even though such arbitrary choices do not prove the complete theoretical consistency of the full theory, this is taken as a proof of concept to investigate phenomenological aspects of the coupling in a cosmological setting. On the other hand, the complete representation of the theory contains non-trivial interacting terms as a result of the metric transformation eqn.~(\ref{sec2:eqn1}) as evidenced, for instance, in eqns. (\ref{sec3:eqn9})-(\ref{sec3:eqn11}).  Notice however that all terms can be classified into two large groups belonging either to conformal or disformal couplings. It means that we can not discard (partially) some terms of $\gamma_{c}$ or $\gamma_{B}$ in (26) since all of them (of $\gamma_{c}$ or $\gamma_{B}$) correspond to one common source. Thereby, supposing that we can apply some guiding principle to the theory, this would constrain the kind of coupling itself and not  each term derived from  it. A more pragmatic strategy to constrain the models studied is, for instance, to find observational evidence in favour of or against some  kind of coupling based on the goodness-of-fit criteria in parameter estimation procedure (see e.g Ref.~\cite{vandeBruck:2016hpz}).

We have also made substantial progresses on the issue of stability of the theory. Concretely, general conditions on the coupling functions to avoid propagation of spurious degrees of freedom in the theory were found. This is translated into the specific models studied as follows:  the free parameters $q$ and $\alpha$ for the conformal cases are unconstrained but for the disformal case we have obtained the ghost-free condition $B_{0}>0$. This is also consistent with the dynamical system constraints (see Table \ref{table2}). Numerical methods to investigate the time evolution of linearized perturbations on a fixed background are however required to verify our preliminary findings. Moreover, the equation of motion for the vector field eqn.~(\ref{sec3:eqn6}) corresponds to a primary constraint, which means that there is no one degree of freedom propagating because it was eliminated. Likewise, we have observed that all our numerical solutions of the background equations are dynamically well-behaved for the explored parameter space. On the other hand, we would like to emphasize that the presence of ghost fields in a theory cannot be determined solely by examining the structure of the equations of motion, as an overall sign in front of the Lagrangian has no influence at all. While the analysis carried out in section \ref{estability}, have helped to identify classical instabilities such as tachyonic ghost and Laplacian instabilities (and a well-posed initial value as well), it does not guarantee the absence of ghosts in the theory. To determine the stability of the theory, a proper Hamiltonian analysis must be carried out to determine whether the Hamiltonian is bounded from below (see e.g \cite{Sbisa:2014pzo,Gumrukcuoglu:2016jbh}), which is not an easy task in curved spacetime \cite{Langlois:2015skt}. If so, this will prevent the propagation of highly excited modes, thereby rendering the theory quantum mechanically stable. This is a crucial aspect that must be assessed for a theoretical consistency of the theory.

We have first investigated the cosmological solutions of the system based on dynamical system techniques to set the stability conditions in terms of the model parameters. Several novel critical points have been found and reported respectively in table (\ref{table1}) for the conformal coupling cases and table (\ref{table2}) for the disformal case, as well as some inferred constraints on the model parameters from purely theoretical grounds. Thus, different types of trajectories can exist in phase space describing the evolution of the universe. This depends essentially on the model parameters that allow the existence of the critical points themselves and the coupling type in consideration. We summarize next the most intriguing solutions. 

The fixed points ($\rm C$) and ($\rm \tilde{F}$), which are saddle points, correspond to a \textit{vector-dark matter scaling solution} for the conformally power law and exponential coupled models, respectively. These kind of solutions are particularly interesting aiming at solving the coincidence problem. The fixed point ($\rm \tilde{D}$), associated to the exponential coupling,  represents an attractor solution so that it can account for the accelerated expansion of the universe. Up to the best of our knowledge the uncoupled solution ($\rm D$), in the exact form proposed here, had not been reported in the literature. Hence, it corresponds to the minimal realization of the model. As to the disformal case, several novel critical points have been obtained as well, apart from those solutions in common with the conformal cases ($\rm A_{\pm}$), ($\rm B_{\pm}$), ($\rm E_{\pm}$), ($\rm D$) and ($\rm S$). The corresponding novel  fixed points are reported in table (\ref{table2}) for two particular choices of the power law parameter $\beta$. In particular $\beta=-1/2$ provides a stable attractor solution ($\rm H$) whose stability is ensured trivially and a saddle point ($\rm G$).
$\beta=-2$ also provides several critical points whose dynamical character does depend on the model parameters. They can be either saddle points (($\rm \tilde{H}$) with $c=3$ and ($\rm \tilde{G}$)) or a stable attractor solution (($\rm \tilde{H}$) with $c=2$), so it is possible to find a region of the parameter space where stability is guaranteed. The latter point is essentially important to drive the current accelerated expansion. We have also identified some solutions where the vector field does not contribute to the energy density in the form of dark energy but it can be present in the dark matter dominated era so that this solution may deviate from the standard dark matter domination solution and, therefore, to leave some observational imprints on the structure formation. 

In addition to the dynamical system analysis, numerical methods have been implemented for completeness to visualize the effects of the coupling parameters on the cosmological dynamics. As a general conclusion, the energy density of dark matter can be lowered or increased depending on the strength on the respective coupling parameter and in some cases, such as the conformally exponential coupling and the disformally coupled models, on their associate signs. Specifically, in the conformally coupled power law model the coupling parameter can affect more significantly the cosmological dynamics during different stages of the evolution of the universe. The same conclusion also applies for the disformally coupled model but with much less distinguishable changes. This suggests that observational data at different redshifts can be used strategically in the future to put constraints on the coupling parameters, in a joint way to the ones derived here from purely theoretical grounds, by implementing standard statistical methods for cosmological parameter estimation. Hence, whether this kind of vector coupled models of dark energy is statistically preferred by observational data or not is a subject that must be investigated in the future to determine their cosmological viability. In particular, coupled dark energy models have shown great potential to solve the Hubble tension, and have been categorized as promising models within $3\sigma$ level to the light of this tension \cite{DiValentino:2021izs}. Coupled (and uncoupled) scalar fields models of dark energy are more mainstream, but we have showed in this paper that coupled vector fields are also appealing at the cosmological background level. We expect thereby to push coupled vector field models of dark energy towards an observational setting by encouraging more people to work in this arena, specifically those working in statistical methods to constrain this class of coupled models with observational data. 


\acknowledgments
We thank anonymous referee for the critical questioning on physical grounds of the theory that help to clarify some important points of this work. G. G acknowledges financial support from Agencia Nacional de Investigaci\'on y Desarrollo ANID through the FONDECYT postdoctoral Grant No. 3210417. We  would like to thank Yeinzon Rodr\'iguez, Carlos M. Nieto and Cl\'ement Stahl for insightful discussions and valuable comments, and also to Alexander Gallego Cadavid for careful reading of the manuscript. Special thanks go to Norman Cruz and Guillermo Palma for all the valued support during my stay in Chile.

\appendix*

\section{Disformal transformations}
We report here some explicit calculations that are needed to go from one frame to another:
\begin{equation}\begin{array}{l}
\frac{\partial \bar{g}_{\mu \nu}}{\partial g_{\alpha \beta}}=C \delta_{\mu}^{\alpha} \delta_{\nu}^{\beta}+\frac{1}{2} A^{\alpha} A^{\beta}\left(C_{X} g_{\mu \nu}+B_{X} A_{\mu} A_{\nu}\right), \\
\frac{\partial g_{\mu \nu}}{\partial \bar{g}_{\alpha \beta}}=\frac{1}{C}\left[\delta_{\mu}^{\alpha} \delta_{\nu}^{\beta}-\frac{1}{2} D A^{\alpha} A^{\beta} \left(C_{X} g_{\mu \nu}+B_{X} A_{\mu} A_{\nu}\right)\right], \\
\frac{\partial \bar{g}_{\alpha \beta}}{\partial A_{\mu}}=B\left(\delta_{\alpha}^{\mu} A_{\beta}+\delta_{\beta}^{\mu} A_{\alpha}\right)-(C_{X} g_{\alpha \beta}+B_{X} A_{\alpha} A_{\beta}) A^{\mu},
\end{array}
\end{equation}\label{sec2:eqn15}
with $C\neq0$. We remind that the quantity $D$ is defined just after eqn.~(\ref{sec3:eqn1}). The inverse map between the two metrics $\bar{g}_{\mu\nu}\to g_{\mu\nu}$ exists around any point provided that the Jacobian has no null-eigenvalues. In addition, the well defined inverse metric needs to be non-singular, causal and preserve Lorentz signature. Also, the inverse transformation requires to keep the same functional dependence for each metric either purely conformal or disformal. So, the transformation only exists when these regularity conditions are met.

\bibliography{apssamp}
\bibliographystyle{apsrev4-1}
\end{document}